\documentclass[a4paper, 10pt]{article}
\usepackage{amsmath,amssymb,amsthm,graphicx,braket,multirow,authblk,amsthm,dcolumn,latexsym,subcaption,mathtools,cite}
\def\oper{{\mathchoice{\rm 1\mskip-4mu l}{\rm 1\mskip-4mu l}
{\rm 1\mskip-4.5mu l}{\rm 1\mskip-5mu l}}}
\usepackage[left=2cm,right=2cm,top=2cm,bottom=2cm]{geometry}
\usepackage[utf8]{inputenc}
\usepackage{hyperref}
\usepackage{url}
\usepackage{amssymb}
\usepackage{bm}
\usepackage{amsfonts}
\usepackage{indentfirst}
\usepackage{enumitem}

\usepackage[font=small]{caption}
\usepackage{cite}
\usepackage{xcolor}

\newcommand{\tr}{\textup{Tr}}

\def\<{\langle}
\def\>{\rangle}
\newcommand{\idm}{\mathbf{1}}

\newcommand{\cK}{{\cal K}}
\newcommand{\cL}{{\cal L}}

\newcommand{\cD}{{\cal D}}

\newcommand{\cT}{{\cal T}}

\renewcommand{\ket}[1]{\left| #1 \>}
\renewcommand{\bra}[1]{\< #1 \right|}

\newtheorem{Def}{Definition}
\newtheorem{Prop}{Proposition}
\newtheorem{Theorem}{Theorem}

\theoremstyle{definition}
\newtheorem{Remark}{Remark}

\begin{document}

\title{{\bf Random Lindblad operators obeying  detailed balance }}
%condition}}
	
	\author[$\hspace{0cm}$]{Wojciech Tarnowski$^{1,}$\footnote{wojciech.tarnowski@doctoral.uj.edu.pl}}
	\affil[$1$]{\small Institute of Theoretical Physics, Jagiellonian University,
	ul. {\L}ojasiewicza 11, 30-348 Cracow, Poland}

	\author[$\hspace{0cm}$]{Dariusz Chru\'sci\'nski$^{2,}$\footnote{darch@fizyka.umk.pl}}
	\affil[$2$]{\small Institute of Physics, Faculty of Physics, Astronomy and Informatics, Nicolaus Copernicus University, Grudzi\c adzka 5/7, 87-100 Toru\'n, Poland}
	
\author[$\hspace{0cm}$]{Sergey Denisov$^{3,4,}$\footnote{sergiyde@oslomet.no}}
	\affil[$3$]{\small  Department of Computer Science, Oslo Metropolitan University, N-0130 Oslo, Norway}
    \affil[$4$]{\small  NordSTAR - Nordic Center for Sustainable and Trustworthy AI Research, Pilestredet 52, N-0166, Oslo, Norway}
\author[$\hspace{0cm}$]{Karol \.Zyczkowski$^{1,5,}$\footnote{karol.zyczkowski@uj.edu.pl}}
	\affil[$5$]{\small Center for Theoretical Physics, al. Lotnik{\'o}w 32, 02-668 Warsaw, Poland}

	\maketitle
	\vspace{-0.5cm}
	
	\begin{abstract}
We introduce  different ensembles of random Lindblad operators $\cL$,
which satisfy quantum detailed balance condition with respect to
 the given stationary state $\sigma$ 
 of size $N$, and investigate their spectral properties. Such operators are  known as `Davies generators' and their eigenvalues are real; however, their spectral densities depend on $\sigma$.
We propose different structured ensembles of random matrices, which allow us to
tackle the problem analytically in the extreme
cases of Davies generators corresponding to
random $\sigma$ with a non-degenerate spectrum
or the maximally mixed stationary state, $\sigma = \idm /N$.
Interestingly, in the latter case the density can be reasonably well approximated
by integrating out the imaginary component of the spectral density 
characteristic to the ensemble of random unconstrained  Lindblad operators. 
%with complex spectrum. 
The case of asymptotic states with partially degenerated spectra is also addressed.
Finally, we demonstrate that similar universal properties hold for the detailed balance-obeying Kolmogorov generators obtained by applying superdecoherence to an ensemble of random Davies generators. 
In this way we construct an ensemble
of random classical generators with imposed detailed balance condition.
	\end{abstract}
	
%\date{\today}
\maketitle

\vskip 0.7cm

{\sl Dedicated to the memory of G{\"o}ran Lindblad, (1940-2022)}

\section{Introduction}

In classical physics the principle of detailed balance  essentially states that, for a system at equilibrium, the rate of the elementary transition from state $a$ to state $b$, $a \to b$, is the same as the rate of the  reverse transition $b \to a$ \cite{Kampen}. 
This  principle perfectly works in chemistry and biology where  transition $a \to b$ is interpreted as a particular chemical reaction or transition between populations of two species, $a$ and $b$~\cite{Kampen}. Classical detailed balance is a property of the corresponding generator of the Markovian evolution of probability vector $\mathbf{p}=(p_1,\ldots,p_N)$ which is  governed by the so-called Pauli rate equation,

\begin{equation}
\label{Pauli-1}
  \dot{p}_i = \sum_{j=1}^N \Big( W_{ij} p_j - W_{ji} p_i \Big) ,
\end{equation}
where non-negative numbers $W_{ij}$ are interpreted as transition rates.
If $\mathbf{p}$ is a stationary state,
 then the detailed balance condition w.r.t. $\mathbf{p}$ means that
\begin{equation}
\label{CDB}
  W_{ij} p_j = W_{ji} p_i ,
\end{equation}
for any pair $i,j$. Condition (\ref{CDB}) is just a mathematical representation of the principle that if the system is in the stationary state $\mathbf{p}$, then the probabilities of transition $i \to j$ and $j \to i$ are the same.

In the quantum case, Markovian evolution is governed by the celebrated Gorini-Kossakowski-Lindblad-Sudarshan (GKLS) master equation $\dot{\rho}_t = \mathcal{L}(\rho_t)$, with the corresponding GKLS-generator\footnote{Henceforth, we will address these objects, in a interchangeable way, as `Lindblad operators' and `GSKL-generators'} having the well-known structure \cite{GKS,L,40-GKLS}:

\begin{equation}
\label{GKLS}
  \mathcal{L}(\rho) = -i[H,\rho] + \sum_{\alpha=1}^{N^2-1}  L_{\alpha} \rho L_{\alpha}^\dagger -  \frac 12 \{ L_{\alpha}^\dagger L_{\alpha},\rho\},
\end{equation}
where the Hermitian operator $H$ stands for an effective system's Hamiltonian and $L_{\alpha}$ are  jump (or noise) operators.

Classical detailed balance condition (\ref{CDB}) was generalized to quantum Markovian semigroups in Ref.~\cite{Alicki-DB,Gorini-DB,Gorini-DB2}  (see also Ref.~\cite{Streater}). If $\sigma$ is a stationary state (meaning that $\mathcal{L}(\sigma)=0$), then $\mathcal{L}$ satisfies quantum detailed balance condition if its dissipative part in the Heisenberg picture is Hermitian w.r.t. the inner product $(X,Y)_\sigma = {\rm Tr}( \sigma X^\dagger Y)$ which reduces to the standard Hilbert-Schmidt inner product if the stationary state is maximally mixed; see Section \ref{SEC-QDB} for more details. In particular, if $|e_k\>$ is an eigen-basis of $\sigma$, then the following {\em classical} transition matrix $W_{ij} = \sum_{\alpha} |\<e_i|L_{\alpha}|e_j\>|^2$ satisfies (\ref{CDB}).

The quantum detailed balance (QDB) condition is a characteristic property of quantum Markovian semigroup describing the
interaction of a system with an environment at equilibrium~\cite{LS}. In particular, it has been shown by Davies \cite{Davies1,Davies2,Davies3} that quantum Markovian semigroup derived in the weak coupling limit satisfies QDB condition w.r.t. stationary state  $\sigma$. Assuming the standard form of the system-environment Hamiltonian \cite{Open1,Open2,Open3,ALICKI},
\begin{equation}
\label{HHS}
  H = H_S \otimes \oper_E + \oper_S \otimes H_E + H_{\rm int},
\end{equation}
and thermal initial state of the environment, $\rho_E = e^{-\beta H_E}/Z_E$, Davies proved that, in the weak coupling limit, $\sigma = \sigma_\beta:=  e^{-\beta H_S}/Z_S$, the corresponding generator (\ref{GKLS}) satisfies QDB condition w.r.t. $\sigma_\beta$. Moreover, in this case unitary part generated by $H_S$ (corrected by the Lamb shift) and dissipative part of $\mathcal{L}$ commute which implies that, during the evolution of the density operator $\rho_t$, the diagonal and off-diagonal elements of the density matrix decouple. Many open quantum systems studied in the literature \cite{Open1,Open2,Open3,ALICKI} fit into this class. The corresponding  generators are often called `Davies generators' (or `Davies GKLS-generators')  and the corresponding dynamical map $\Lambda_t = e^{t \mathcal{L}}$ is referred to as a `Davies map'\cite{Roga}.

The quantum detailed balance condition plays an important role in quantum thermodynamics of open quantum systems~ \cite{ALICKI,LS} (see also recent works~ \cite{QT1,QT2,QT3,QT4,QT5,QT6,QT7}). In particular, assuming that the QDB holds, the Second Law of thermodynamics stating that  the entropy production rate is never negative can be formulated as~\cite{LS}:

\begin{equation}\label{lab3}
  \frac{d}{dt} S(e^{t\mathcal{L}}(\rho) |\!| \sigma_\beta) \leq 0 ,
\end{equation}
where $S(\rho|\!|\sigma)$ is the relative entropy and $\rho$ is an arbitrary state of the system of size $N$.

Another interesting property of the GKLS-generators  obeying  the DB condition  was observed in Ref.~\cite{GEN}. Let $\ell_\alpha$
denote (in general complex) eigenvalues of $\mathcal{L}$.
One has $\mathcal{L}(\sigma)=0$ and it is well known that ${\rm Re}\, \ell_\alpha \leq 0$. Hence one defines (non-negative) relaxation rates $\Gamma_{\alpha} = - {\rm Re}\, \ell_{\alpha}$ $(\alpha=1,\ldots,N^2-1$). Actually, $\Gamma_{\alpha}$ are measurable quantities. It was conjectured~\cite{GEN} that
\begin{equation}\label{GGG}
  \Gamma_{\alpha} \leq  \frac 1N (\Gamma_1 + \ldots +\Gamma_{N^2-1}) .
\end{equation}
Interestingly, this inequality was proven for $\cL$ satisfying quantum detailed balance~\cite{GEN} (see also Ref.~\cite{PR} for a recent review).  
All that hints that GKLS-generators satisfying the detailed balance condition enjoy many interesting properties, both from mathematical and physical points of view.
In particular, spectra of such operators might display some universal features. Note, that since dissipative part of $\mathcal{L}$ is Hermitian w.r.t. inner product $(A,B )_{\sigma}= {\rm Tr}(\sigma A^\dagger B)$, it possesses real spectrum. Hence, purely dissipative Lindblad operators satisfying detailed balance condition, have purely real spectra. Obviously, the same applies to Kolmogorov operators obeying classical detailed balance.

Recently, we analyzed
spectral properties of random Lindblad operators $\cal L$~\cite{Lemon,PRE}
as well as classical Kolmogorov operators $\cal K$~\cite{PRE}.
In particular, we have shown that purely dissipative random Lindblad operators exhibit universal spectral properties, that is, properly rescaled complex eigenvalues form the universal {\it lemon}-like shape on the complex plane (see also recent works on random Lindblad operators~ \cite{random1,random1a,random2,random3}). Such approach enables one to study  the properties of a typical quantum system evolving under the action of Markovian semigroup. Random Matrix Theory (RMT) \cite{Mehta} provides appropriate tools to deal with this problem. RMT has developed into a field with many important applications in physics and mathematics; see  Ref.~\cite{RMT1} and a collection of papers in Ref.~\cite{RMT-JPA}. Starting from a seminal monograph by Haake \cite{Haake} (see also Ref.~\cite{Haake+}), random matrices became a key theoretical tool to study classical and quantum chaos~\cite{chaos1,chaos2}.
For recent application of RMT to open quantum system see, e.g., Ref.~\cite{chaos-open}.

In Ref.~\cite{Lemon} a simple random matrix (RM) model, which reproduces the lemon-like shape of the bulk of complex eigenvalues of random Lindblad operators and provides analytical expression for the boundary of the lemon, was proposed.
 Moreover, using  
free probability tools,
 we constructed structured ensembles of random matrices
 which allow us to describe  the transition from the well-known Girko disc~\cite{Gi84},
 characteristic of random operations \cite{BCSZ09,KNPPZ21},
 to the lemon-like distribution typical of GKLS-generators.

  The classical counterpart, that is
an ensemble of  random Kolmogorov generators, was
analyzed first by Timm \cite{Timm1}
and further discussed by Bordenave, Caputo, and Chafai~\cite{BordenaveChafai}.
It turns out that  the properly rescaled complex eigenvalues of $\cal K$ form the universal {\it spindle}-like shape on the complex plane~\cite{Timm1,BordenaveChafai}.
Interestingly, these two types of operators, Lindblad and Kolmogorov ones, can be related by {\it  superdecoherence}~\cite{PRE}. Superdecoherence is a particular  example of a {\it supermap}, i.e. a linear map which sends a quantum channels into a  quantum channel~
\cite{CDP08,Zy08,Gour}. In analogy with the mechanism of decoherence, which transforms a quantum state into a classical one (w.r.t. a fixed orthonormal basis in the system's Hilbert space),  superdecoherence sends quantum maps into
classical stochastic matrices~ \cite{KCPZ18,Karol-Kamil},
 while quantum Lindblad operators are transformed into classical  Kolmogorov operators~\cite{PRE}.  By gradually increasing strength of superdecoherence, one can observe how the lemon-like shape of the spectral distribution transforms into the spindle-like shape~\cite{PRE}.

In this paper we extend the analysis to the operators satisfying detailed balance condition.
As in this case the spectrum of the operators is real,
we analyze the density of eigenvalues along the real axis for various assumptions
concerning invariant states. In particular,
we consider most relevant examples of
 random state $\sigma$ with a non-degenerate spectrum, which leads to a Davies generator,
and also the fully degenerated stationary state, $\sigma = \oper /N$.
A class of random Davies generators  for stationary partially degenerated states is also
analyzed.

This paper is organized as follows.
Classical and quantum detailed balance conditions
are recalled in Sections
 \ref{SEC-CDB} and  \ref{SEC-QDB}, respectively. Generic random Lindblad and Kolmogorov operators are reviewed in Section~\ref{sec:Lemon}.
Ensembles of random Lindblad  operators, which satisfy the QDB condition with respect to a steady state $\sigma$ with varying eigenvalue degeneracy, are analyzed in Section \ref{random_detailed}.
 Several random matrix models  which allow us
 to approximate the spectral density of random operators,  observed in numerical simulations, are introduced.
Concluding remarks, including a list of  open problems and some possible directions for future research, are presented in  Section \ref{sec:Conclusions}.
Appendix~\ref{sec:Glossary} contains a concise review of basic RMT ensembles used in this work.
A short discussion of the  Pastur equation~\cite{PasturEquation}, which is essential for the description of eigenvalue densities, is presented in Appendix~\ref{sec:Pastur}. Some of the technical details of the results for the random Lindbladian with partially degenerate steady state are presented in Appendices~\ref{sec:LindbladSpectrum} and \ref{sec:Justification}.

\section{Classical detailed balance condition}
 \label{SEC-CDB} 

The operator generating Pauli master equation~\eqref{Pauli-1} is called Kolmogorov generator and given by
\begin{equation}\label{Kij2}
  {\cal K}_{ij} = W_{ij} - \delta_{ij} \sum_{k=1}^N W_{kj}.
\end{equation}
Any classical generator has at least one stationary state $\mathbf{p}$ such that
${\cal K} \mathbf{p} = 0$.

\begin{Def} Classical operator $\cal K$ satisfies detailed balance condition w.r.t. $\mathbf{p}$ if Eq. (\ref{CDB}) is satisfied
for any pair $i,j$.
\end{Def}
The very condition (\ref{CDB}) is just a mathematical representation of the principle that if the system is in  stationary state $\mathbf{p}$, then the probabilities of transition $i \to j$ and $j \to i$ are the same.  In particular,
 if $\mathbf{p}$ is a thermal state at inverse temperature $\beta$,
 i.e. $p_k = e^{-\beta E_k}$, then

\begin{equation}\label{Wij}
 \frac{W_{ij}}{W_{ji}} = e^{-\beta(E_i - E_j)} .
\end{equation}
Interestingly, the detailed balance condition (\ref{CDB}) can be reformulated as follows: Let us define  in $\mathbb{R}^N$
 an inner product
$(.\, ,.)_{\mathbf{p}}$ with respect to a given state $\mathbf{p}$,

\begin{equation}\label{scala}
  (\mathbf{x},\mathbf{y})_{\mathbf{p}} := \sum_{k=1}^N p_k x_k y_k .
\end{equation}

\begin{Prop} An operator $L$ satisfies the detailed balance condition w.r.t. $\mathbf{p}$ if and only if $L^{\rm T}$ satisfies

\begin{equation}\label{LL}
  (L^{\rm T}\mathbf{x},\mathbf{y})_{\mathbf{p}} = (\mathbf{x},L^{\rm T}\mathbf{y})_{\mathbf{p}} ,
\end{equation}
for any $\mathbf{x},\mathbf{y} \in \mathbb{R}^n$, that is, $L^{\rm T}$ is Hermitian w.r.t. $(. \, ,. )_{\mathbf{p}}$.
\end{Prop}
Equivalently, defining  diagonal matrix $P := {\rm Diag}[p_1, \ldots,p_n]$,
 condition (\ref{CDB}) can be rewritten as 

\begin{equation}\label{WPPW}
 W P = P W^{\rm T} .
\end{equation}
In particular, if $\mathbf{p}$ is maximally mixed, then the transition matrix $W^{\rm T}=W$, i.e. $W$ is symmetric. Assuming that $\mathbf{p}$ is faithful, i.e. $p_k > 0$, any transition matrix $W$ satisfying
condition
 (\ref{WPPW}) may be constructed as 

\begin{equation}\label{WSP}
  W = S P^{-1} ,
\end{equation}
where $S$ is an arbitrary (real) symmetric matrix.

\section{Quantum detailed balance}
 \label{SEC-QDB}

Classical detailed balance condition was generalized for quantum Markovian semigroups
 \cite{Alicki-DB,Gorini-DB,Gorini-DB2}.
 Let us first recall the standard Hilbert-Schmidt product of two operators,
  $(X,Y)_{\rm HS}={\rm Tr} X^{\dagger}Y$.
  It will be also convenient to distinguish fixed faithful quantum state $\sigma >0$
 and to introduce a scalar product of any two operators,
 $(X,Y)_\sigma := {\rm Tr}(\sigma X^\dagger Y)$.
 In analogy to the scalar product (\ref{scala}),
 %determined by a classical state $p^{\rm s}$,
 for the maximally mixed state $\sigma=\oper/N$,
 the inner product $(X,Y)_\sigma$ reduces to the standard construction of Hilbert-Schmidt product,  i.e. $(X,Y)_\sigma = \frac 1N (X,Y)_{\rm HS}$. 
These notions allow one to express the quantum condition of detailed balance
 \cite{Alicki-DB,Gorini-DB,Gorini-DB2}.

\begin{Def} A GKLS generator satisfies quantum detailed balance condition
with respect to a given stationary
state $\sigma$   if there exists a representation

\begin{equation}\label{}
  \mathcal{L}(\rho) = -i[H,\rho] + \mathcal{L}_D(\rho) ,
\end{equation}
such that $[H,\sigma]=0$, and the  dissipative part $\mathcal{L}_D$ satisfies

\begin{equation}\label{QDB}
  (\mathcal{L}_D^\ddag(X),Y)_\sigma = (X,\mathcal{L}_D^\ddag(Y))_\sigma ,
\end{equation}
where %$(X,Y)_{\rm s} = {\rm Tr}(\rho^{\rm s} X^\dagger Y)$, and
$\Phi^\ddag$ denotes a dual map (in the Heisenberg picture)
defined via $(\Phi^\ddag(X),Y)_{\rm HS} = (X,\Phi(Y))_{\rm HS}$.
\end{Def}
Note, that if
\begin{equation}\label{eq:Lchannel}
  \mathcal{L}(\rho) = -i[H,\rho] + \Phi(\rho) - \frac 12 \{\Phi^\ddag(\oper),\rho\} ,
\end{equation}
for some completely positive map $\Phi$, then its dual is defined as follows

\begin{equation}\label{}
  \mathcal{L}^\ddag(X) = i[H,X] + \Phi^\ddag(X) - \frac 12 \{\Phi^\ddag(\oper),X\} .
\end{equation}
Quantum detailed balance condition (\ref{QDB}) reduces to $[\Phi^\dagger(\oper),\sigma]=0$ together with the following condition

\begin{equation}\label{QDB-2}
  (\Phi^\ddag(X),Y)_\sigma = (X,\Phi^\ddag(Y))_\sigma ,
\end{equation}
that is, the map $\Phi^\ddag$ is Hermitian w.r.t. $(.\, , . )_\sigma$.
If the stationary state $\sigma$ is faithful  the above condition may be rewritten as follows
\begin{equation}\label{}
  \Phi^\ddag(X) = \Phi(X \sigma) \sigma^{-1} ,
\end{equation}
or, equivalently

\begin{equation}\label{}
  \Phi^\ddag(X) \sigma =  \Phi(X \sigma) ,
\end{equation}
Actually, defining a map (so called modular operator)

\begin{equation}\label{}
  \Delta_\sigma(X) :=  \sigma X \sigma^{-1} ,
\end{equation}
one proves \cite{Alicki-DB} that if $\mathcal{L}$ satisfies quantum detailed balance w.r.t. $\sigma$, then

\begin{equation}\label{}
  [\mathcal{L},\Delta_\sigma]=0 , \ \ \ \ [\mathcal{L}^\ddag,\Delta_\sigma]=0 .
\end{equation}
It is clear that a completely positive map $\Phi$ is a quantum version of  classical transition matrix $W_{ij}$. Let $\{|i\>\}_{i=1}^N$ be an eigen-basis of $\sigma$, i.e. $\sigma = \sum_{i=1}^N p_i |i\>\<i|$ and define

\begin{equation}\label{}
  W_{ij} := \< i|\Phi(|j\>\<j|)|i\> = \< j|\Phi^\ddag(|i\>\<i|)|j\> .
\end{equation}
Then, if $\Phi$ satisfies conditions (\ref{QDB-2}), the transition matrix $W_{ij}$ satisfies the classical condition (\ref{CDB}).

In particular, if  $\sigma= \oper/N$, then the quantum detailed balance condition
  states that $\mathcal{L}_D^\ddag = \mathcal{L}_D$,
   i.e. Schr\"odinger and Heisenberg pictures coincide.
   Hence the dissipative part may be represented as

\begin{equation}\label{}
  \mathcal{L}^\ddag_D(X) = \sum_{\alpha,\beta=1}^{N^2}  K_{\alpha\beta} \Big( F_\alpha X F_\beta - \frac 12 \{ F_\alpha F_\beta,X\} \Big) ,
\end{equation}
with Hermitian operators $F_\alpha$ and real symmetric Kossakowski matrix $K_{\alpha\beta}$.

Let   $E_{ii'} = |i\>\<i'|$, where $|i\>$ defines an eigen-basis of the stationary state,  be an orthonormal basis in the operator space. One has the following representation

\begin{equation}\label{}
  \mathcal{L}_D^\ddag(X) = \sum_{i,i'=1}^N\sum_{j,j'=1}^N K_{ii',jj'} \Big( |i\>\<i'|X|j'\>\<j| - \frac 12 \delta_{i'j'} \{ |i\>\<j|,X\} \Big) ,
\end{equation}
where the Kossakowski matrix $K_{ii',jj'}$ is positive definite. Now, the detailed balance condition (\ref{QDB}) implies the following relation between elements of the Kossakowski matrix~\cite{Gorini-DB2}:

\begin{equation}\label{CC}
  K_{ii',jj'} p_j = K_{j'j,i'i} p_{j'} .
\end{equation}
Note that condition (\ref{CC}) implies

\begin{equation}\label{CC-2}
  K_{ii',jj'} q_{jj'} = K_{ii',jj'} q_{ii'} ,
\end{equation}
where  $q_{mn} := p_m/p_n$. The QDB is a highly restrictive condition, since if $ q_{ii'} \neq q_{jj'}$, the corresponding elements $K_{ii',jj'}$ must vanish. In particular, in a typical case of non-degenerate $\sigma$ satisfying
\begin{equation}
   q_{mn} = q_{m'n'} \ \ \Longleftrightarrow \ \ (m,n) = (m',n') ,  \label{eq:nondegen_cond}
\end{equation}
the only non-vanishing elements of $K$ are the following
\begin{equation}\label{}
  W_{ij} = K_{ij,ij} , \ \ \   D_{ij} = K_{ii,jj}, \ \ \ i,j=1,\dots, N.
\end{equation}
Note, that  $W_{ij} \geq 0$, and $D_{ij} = D_{ji}$.

\begin{Theorem} \label{Thm:Lindblad_nondegen}
The Lindblad generator satisfying QDB condition with respect to a steady state that satisfies \eqref{eq:nondegen_cond} reads
\begin{equation}
\mathcal{L}_{\rm Dav}(\rho) = \Phi(\rho) - \frac 12 \{\Phi^\ddag(\oper),\rho\} \label{eq:Lindblad_nondegen}
\end{equation}
with
\begin{equation}\label{}
  \Phi(\rho) = \sum_{i\neq j} W_{ij}  |i\>\<j| \rho |j\>\<i| + \sum_{i,j} D_{ij} |i\>\<i| \rho |j\>\<j| , \label{eq:Drepr}
\end{equation}
and the transition matrix $W_{ij}$ satisfies the classical detailed balance condition $W_{ij} p_j = W_{ji} p_i$.
\end{Theorem}
The above generator is usually called the `Davies generator'~\cite{Davies1}.
Davies generator, apart from the classical part defined in terms of $W_{ij}$, contains a purely quantum part, defined in terms of $D_{ij}$, which is responsible for the decoherence process.

The second extreme scenario corresponds to the maximally mixed state,
 $\sigma =\oper/N$,
 for which condition \eqref{CC} reduces to
% In this case one has the following constraints for the Hermitian matrix $K_{ii',jj'}$
\begin{equation}\label{CC2}
  K_{ii',jj'}  = K_{j'j,i'i},
\end{equation}
allowing for all elements of $K$ to be non-zero. The matrix $K_{ii',jj'}$ is Hermitian, however~\eqref{CC2} does not imply symmetry. Nevertheless, when Lindblad operator is represented in the Hermitian basis $F_\alpha$, QDB implies $K_{\alpha\beta}=K_{\beta\alpha}$.

\begin{Remark} Interestingly, the inner product (\ref{QDB-2}) can be generalized to the following one-parameter family of products~\cite{Fagnola,Fagnola2,Carlen,KMS-S}

\begin{equation}\label{}
  (X,Y)_s := {\rm Tr}(\sigma^s X^\dagger \sigma^{1-s} Y) ,
\end{equation}
for $s\in [0,1]$. Note that (\ref{QDB-2}) is recovered for $s=1$ and it is often called `GNS product'. For $s=\frac 12$ one obtains so called `KMS product',
  
\begin{equation} \label{KMS}
  (X,Y)_{\rm KMS} := (X,Y)_{1/2} = {\rm Tr}(\sigma^{1/2} X^\dagger \sigma^{1/2} Y) .
\end{equation}
It turns out~\cite{Fagnola,Fagnola2} that if $\mathcal{L}_D^\ddag$ is self-adjoint w.r.t. $(X,Y)_1$, then $\mathcal{L}_D^\ddag$ is also self-adjoint for any $s\neq \frac 12$. The symmetric case corresponding to $s=\frac 12$ was analyzed in Ref.~\cite{Fagnola2}. 
\end{Remark}

%%%%%%%%%%%%%%%%%

\section{Spectral properties of random Lindblad and Kolmogorov operators without the detailed balance}
\label{sec:Lemon}

Let us briefly recall the main properties of random Lindblad operators\cite{Lemon,PRE} and random Kolmogorov operators \cite{Timm1,BordenaveChafai,PRE}. Assuming that the Hamiltonian part vanishes, a Lindblad operator has only a dissipative part fully controlled by a completely positive map $\Phi$, see Eq.~(\ref{eq:Lchannel}).  Fixing  orthonormal basis $\{F_\alpha\}_{\alpha=1}^{N^2}$, it can be represented via $\Phi(\rho) = \sum_{\alpha,\beta=1}^{N^2} K_{\alpha\beta}F_\alpha \rho F_\beta^\dagger$, where the so-called Kossakowski matrix $K$ (with matrix elements $K_{\alpha\beta}$) is positive definite. In what follows we use the following normalization ${\rm Tr}K=N$ which is equivalent to

\begin{equation}\label{normalization}
  {\rm Tr}\,\Phi(\oper) = {\rm Tr}\,\Phi^\ddag(\oper) = N .
\end{equation}
The spectrum of $\mathcal{L}$ coincides with the spectrum of the corresponding super-operator

\begin{equation}
  \widehat{\mathcal{L}} = \widehat{\Phi} - \frac{1}{2} ( \Phi^\ddag(\oper) \otimes \oper + \oper \otimes  \overline{\Phi^\ddag(\oper)}) ,
\end{equation}
where $\widehat{\Phi} = \sum_{\alpha,\beta}K_{\alpha\beta} F_{\alpha} \otimes \overline{F}_{\beta}$ is obtained by vectorization.
Now, a random operator $\mathcal{L}$ corresponds to a random CP map $\Phi$ and hence to a random Kossakowski matrix $K_{\alpha\beta}$  which can be sampled as a Wishart matrix

\begin{equation}\label{K-GG}
  K =  N \frac{GG^\dagger}{{\rm Tr}\,GG^\dagger} ,
\end{equation}
 where $G$ is a complex $N^2 \times N^2$ square Ginibre matrix.  It turns out \cite{Lemon,PRE} that  the rescaled operator $\widehat{\mathcal{L}}'=N(\widehat{\mathcal{L}} + \oper \otimes \oper)$ displays in the large $N$ limit the universal lemon-shape distribution of eigenvalues; see Fig.~1a.

This universal shape can be  explained by the following  random matrix (RM) model ~\cite{Lemon,PRE}. By representing  positive matrix $K_{\alpha\beta}$ as $K =  G^\dagger G$, with $G$ being a complex Ginibre matrix with the variance

\begin{equation}\label{}
  \< G_{\alpha\mu}\overline{G}_{\beta\nu} \> = \frac{1}{N^3} \delta_{\alpha\beta}\delta_{\mu\nu} ,
\end{equation}
one obtains $\< {\rm Tr} K\> = N$ (that is, we require that ${\rm Tr}K$ equals $N$ only on average). By introducing jump operators,

\begin{equation}\label{eq:JumpLemon}
  L_\alpha = \sum_{\mu=1}^{N^2} G_{\alpha\mu} F_\mu ,
\end{equation}
one obtains the  diagonal Kraus representation

\begin{equation}\label{}
  {\Phi} = \sum_{\alpha=1}^{N^2} L_\alpha \rho {L^\dagger_\alpha} ,
\end{equation}
where the (random) Kraus operators satisfy

\begin{equation}\label{LL}
  \< {\rm Tr} \, L_\alpha L^\dagger_\beta \> = \frac 1N \delta_{\alpha\beta} .
\end{equation}
Now, since the entries of $L_\alpha$ are i.i.d. Gaussian variables in the large $N$ limit, (complex) eigenvalues  of $L_\alpha$ uniformly cover the disk of radius $r=1/N$ on the complex plane.
The corresponding super-operator $\widehat{\mathcal{L}}$ has the  form

\begin{equation}\label{}
  \widehat{\mathcal{L}} = \sum_{\alpha=1}^{N^2}  L_\alpha \otimes \overline{L_\alpha}  - \oper \otimes \oper - \frac 12 (X \otimes \oper + \oper \otimes \overline{X}) ,
\end{equation}
where Hermitian operator $X$ is defined by

\begin{equation}\label{X-matrix}
  X = \Phi^\ddag(\oper) - \oper  = \sum_{\alpha=1}^{N^2}  L^\dagger_\alpha L_\alpha - \oper .
\end{equation}
Hence the spectrum of $\widehat{\mathcal{L}}$ is controlled by the spectra of $\widehat{\Phi}$ and $X \otimes \oper + \oper \otimes \overline{X}$. Note that eigenvalues of $L_\alpha \otimes \overline{L_\alpha}$  in the large $N$ limit are uniformly distributed  on a disk of radius $1/N^2$. The super-operator $\widehat{\Phi}$ is a sum
of $N^2$ independent matrices $L_\alpha \otimes \overline{L_\alpha}$. Hence, in the large $N$ limit (according to
the central limit theorem for non-Hermitian matrices~\cite{Tao}), its spectral density is uniform on the disk of radius $1/N$. It is, therefore, clear that in the large $N$ limit, $\widehat{\Phi}$ can be modeled as a Ginibre matrix with the spectral radius $1/N$. Actually, since the map $\Phi$ preserves Hermiticity, $\widehat{\Phi}$ can be modeled as a real Ginibre matrix $G_R$. For the second term, $X \otimes \oper + \oper \otimes \overline{X}$, let us observe that due to (\ref{X-matrix}) one has $X = \sum_{\alpha=1}^{N^2}X_\alpha$, with $X_\alpha  =  L_\alpha^\dagger L_\alpha - \frac{1}{N^2}\oper$, i.e. $X_\alpha$ is a Wishart matrix $L_\alpha^\dagger L_\alpha$ shifted by $1/N^2$. Note that the distribution of eigenvalues of $X_\alpha$  has zero mean and variance $ 1/N^2$.
Applying now the free central limit theorem \cite{Voiculescu} to  random matrix $X$, which constitutes a sum of $N^2$ matrices $X_\alpha$, one finds that the spectral density of $X$ is defined by the Wigner semicircle supported on $[-\frac 2N,\frac 2N]$, 

\begin{equation}\label{}
  \rho_X(x) =\frac{N^2}{2\pi} \sqrt{ \frac{4}{N^2} - x^2} .
\end{equation}
It is, therefore, evident that $\widehat{\mathcal{L}}' := N(\widehat{\mathcal{L}} + \oper_N \otimes \oper_N)$ could be approximated by the following RM model \cite{Lemon,PRE}

\begin{equation}\label{RMM-1}
  \widehat{\mathcal{L}}' \approx G_R + \frac{1}{2}(\textup{GOE}_N \otimes \oper_N + \oper_N \otimes \textup{GOE}_N) ,
\end{equation}
where in both terms we have the same GOE matrix, the spectral density of which is given (at the large $N$ limit) by the Wigner semicircle on $[-2,2]$,
\begin{equation}\label{Wigner-C}
  \rho_{\rm GOE}(x) = \frac{1}{2\pi} \sqrt{4-x^2} .
\end{equation}
The spectral density of the above RM model is given by a free convolution of the unit Girko disk with a (classical) convolution of two Wigner semicircles of two (rescaled by 1/2) GOE matrices (see Refs.~ \cite{Lemon,PRE} for more details). Interestingly, this simple RM model reproduces lemon-like shape of the spectral distribution of random Lindblad operators. The boundary of the lemon is characterized by the solution of the following equation \cite{Lemon,PRE}:

\begin{equation}\label{boundary}
{\rm Im}[z + G(z)] = 0,
\end{equation}
with

\begin{equation}\label{eq:GLindblad}
G_{\cal L}(z) = 2z - \frac{2z}{3\pi} \left[ (4 + z^2 )E\left( \frac{4}{z^2}\right) +  (4 - z^2 )K\left( \frac{4}{z^2}\right) \right] ,
\end{equation}
where $K(z)$ and $E(z)$ are the complete elliptic integrals of the first and second kind, respectively (cf. \cite{Lemon,PRE} for details). Additionally, the density of complex and real eigenvalues can be calculated, however the resulting formulas are rather involved~\cite{PRE}. Interestingly, the density is constant in the imaginary direction.

\begin{figure}\begin{center}
\includegraphics[width=0.75\textwidth]{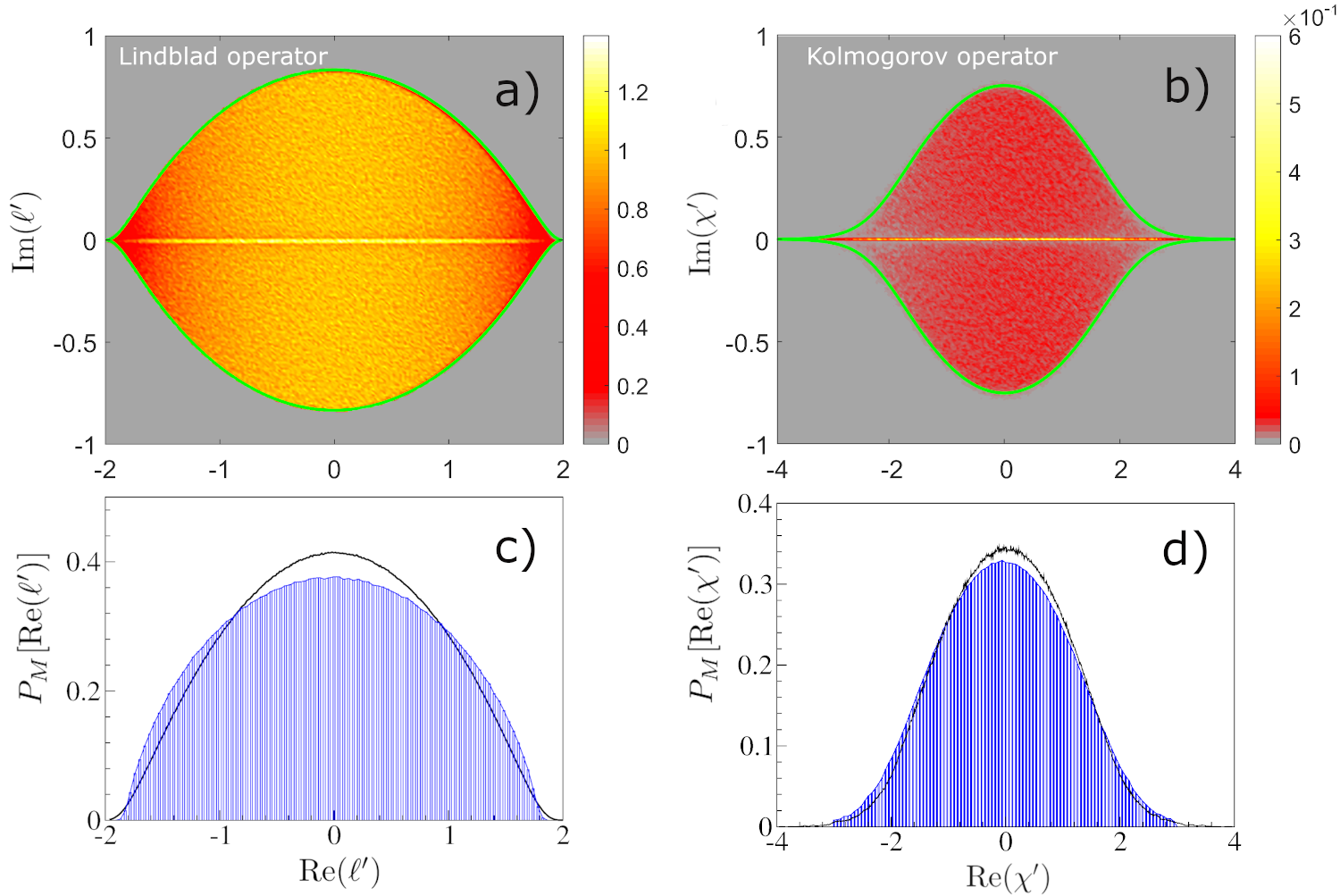}
\end{center}
\caption{a) Spectral density 
	$P[\mathrm{Re}(\ell'),\mathrm{Im}(\ell')]$
	of the rescaled eigenvalues,	$\ell'= N (\ell +1)$,
	from  the spectrum
	of random purely dissipative Lindblad operators  ${\cal L}$  for  $N=250$. Bright green contour is the spectral border, Eqs.~(\ref{boundary},\ref{eq:GLindblad}). The density was sampled with $10$  realizations.
 b)  Probability density functions
$P[\mathrm{Re}(\chi'),\mathrm{Im}(\chi')]$
of  eigenvalues	$\chi'= \sqrt{N} (\chi'+1)$,
of Kolmogorov operators  $\cal{K}$ obtained by decoherifying the ensemble of random
Lindblad operators. The density was sampled with $10^2$~realizations. Bright green contour is the spectral border, Eqs.~(\ref{boundary},\ref{eq:GKolmogorov}).
c) Marginal density of real part of eigenvalues, $\mathrm{Re}(\ell')$, of random Lindblad operators (black line) and the spectral density of a random Lindblad operator satisfying detailed balance conditions w.r.t. state $\sigma = \oper /N$ (blue bars). c) Marginal density of real part of eigenvalues, $\mathrm{Re}(\ell')$,  of random Kolmogorov operators (black line) the spectral density of a random Kolmogorov operator satisfying detailed balance conditions w.r.t. state $\mathbf{p} = (1/N,1/N,...,1/N)$ (blue bars).
}
\label{fig:Lemon}
\end{figure}

Similar analysis can be performed for random Kolmogorov operators~ \cite{Timm1,BordenaveChafai,PRE}. One has $\mathcal{K}_{ij} = W_{ij} - \delta_{ij} W_j$, with $W_j = \sum_k W_{kj}$, and $W_{ij} \geq 0$. To generate $\cK$, the elements of $W$ are i.i.d sampled from a distribution supported on a positive half-line. We choose $W_{ij}=|z_{ij}|^2$, where $z_{ij}$ are i.i.d. complex Gaussian with zero mean and $N^{-1/2}$ variance. The rescaled operator $\cK' = \sqrt{N}(\cK + \oper)$, in the large $N$ limit, exhibits universal spindle-shaped distribution of its eigenvalues, see Fig.~\ref{fig:Lemon}b. This distribution can be  obtained  with the following  simple RM model. Matrix $W$ is modeled by the real Ginibre matrix $G_R$ with the spectral radius 1. By the central limit theorem the diagonal elements $W_j$ can be approximated by Gaussians. One arrives at the following model

\begin{equation}\label{K'}
  \mathcal{K}' \approx G_R + \textup{Gauss}_N,
\end{equation}
where $\textup{Gauss}_N$ represents a diagonal matrix whose elements are i.i.d. Gaussians with zero mean and unit variance.
This model  implies that the distribution of eigenvalues is governed by a free convolution of  a Girko disk and a real Gaussian distribution. The boundary of the spindle-like shape can be calculated from (\ref{boundary}), where now the function $G(z)$ reads

\begin{equation}\label{eq:GKolmogorov}
  G_{\mathcal{K}}(z) = \sqrt{\frac{\pi}{2}} e^{-z^2} \left[  \mbox{Erfi}\left( \frac{z}{\sqrt{2}}\right) - i {\rm sgn}({\rm Im}\, z) \right] ,
\end{equation}
where $\mbox{Erfi}(z) = - i{\rm Erf}(iz)$ and ${\rm Erf}(z) = \frac{2}{\sqrt{\pi}} \int_0^z e^{-t^2} dt$ is the
error function.

\section{Random Lindblad operators satisfying the detailed balance condition}
\label{random_detailed}

In this section we study a family of ensembles of random Lindblad operators satisfying the QDB conditions.
Such operators are parameterized by an invariant state $\sigma$ of size $N$, which can be chosen arbitrarily. Degeneracy of eigenvalues of $\sigma$, more precisely, relation~\eqref{CC-2},  determines which elements of the corresponding Kossakowski matrix are allowed to be non-zero. We start the analysis with two extreme cases, namely a fully degenerate density matrix corresponding to maximally mixed steady state $\sigma = \oper /N$, and a density matrix with no degeneracies, which is a typical scenario if that matrix is drawn at random. Then, we consider a scenario in which the degeneracy of $\sigma$ gradually increases, thus interpolating between the extreme cases.

\subsection{The detailed balance with respect to maximally mixed state $\sigma=\oper/N$}
\label{sec:Davies_degen}

If $\sigma = \oper/N$, then $\mathcal{L}$ satisfies QDB w.r.t. $\sigma$ if and only if $\Phi$ is self-dual, i.e. $  \Phi = \Phi^\ddag$. We denote such Lindblad operator as $\cL_{\rm mm}$. In particular, if the basis $F_\alpha$ consists of Hermitian operators, then $\Phi = \Phi^\ddag$ if and only if $K_{\alpha\beta}= K_{\beta\alpha}$, and hence the Kossakowski matrix $K$ is real symmetric.  It can be sampled according to Eq.~\eqref{K-GG}, where now $G$ is a  $N^2 \times N^2$ real Ginibre matrix. The corresponding jump operators in Eq.~\eqref{eq:JumpLemon} are now Hermitian, which is the main technical consequence of QDB. This provides an alternative and more efficient method of sampling random Lindblad operators by generating jump operators as i.i.d. GUE matrices of size $N$ normalized to $\<\tr L_{\alpha}^2\>=\frac{1}{N}$.

\begin{Remark} In the canonical representation of GKLS- generator, Eq.~ \eqref{GKLS},  jump operators $L_\alpha$ are traceless. However, one can easily check that if $L_\alpha$ is Hermitian, then
\begin{equation}\label{}
  L_{\alpha} \rho L_{\alpha} - \frac 12 \{ L_{\alpha} L_{\alpha} ,\rho \} = \widetilde{L}_{\alpha} \rho \widetilde{L}_{\alpha} - \frac 12 \{ \widetilde{L}_{\alpha} \widetilde{L}_{\alpha} ,\rho \} ,
\end{equation}
where $\widetilde{L}_{\alpha} = {L}_{\alpha} - \frac 1N {\rm Tr}L_\alpha \oper$. Therefore, tracelessness of $L_\alpha$ is not essential.
\end{Remark}

The justification of the RM model follows the reasoning presented in Section~\ref{sec:Lemon} with tiny adjustments. Namely, since jump operators are Hermitian, the super-operator $\hat{\Phi}$ is also Hermitian and it can be modeled with a GOE matrix of size $N^2$. Therefore, the rescaled Lindbladian $\hat{\cL}'_{\rm mm} = N(\hat{\cL}_{\rm mm} + \oper \otimes \oper)$ can be approximated with the following RM model:
\begin{equation}
\label{eq:Lindblad_max_mixed}
\hat{\cL}'_{\rm mm} \approx \textup{GOE}_{N^2}^{(1)} + \frac{1}{2}(\textup{GOE}_N^{(2)} \otimes \oper_N + \oper_N \otimes \textup{GOE}_N^{(2)}),
\end{equation}
where GOE matrices with different superscripts are independent.

To calculate the density of the above RM model, we resort to free probability toolbox for Hermitian matrices~\cite{VND,MSp17}. Since the direct treatment of the density is unhandy, one usually resorts to its Stieltjes transform, also known as Green's function
\begin{equation}
G(z) = \left<\frac{1}{N}\tr\, (z-H)^{-1}\right> = \int \frac{\rho(x)}{z-x}dx.
\end{equation}
Once the Green's function is obtained, the density can be then recovered through the Sokhocki-Plemenlj formula
\begin{equation}
\rho(x) = \mp \frac{1}{\pi} \lim_{\epsilon \to 0} G(z\pm i\epsilon).
\end{equation}
In free probability, the problem of addition of random matrices is solved through free convolution~\cite{Voiculescu,VND,MSp17}. In our RM model the spectrum of  matrix $C \otimes \oper_N + \oper_N \otimes C$ is given by a (classical) convolution of two Wigner semicircles. Then, the resulting density undergoes a free convolution with another Wigner semicircle.

\begin{Prop} \label{Prop:GMaxMixed}
The Stjeltjes transform $G(z)$ of $\cL'_{\rm mm}$ satisfies the equation
\begin{equation}
G(z) = G_{\cL}(z-G(z)) 
\label{eq:Pastur_numeric}
\end{equation}
with $G_{\cL}$ given by \eqref{eq:GLindblad}.
\end{Prop}
The proof is presented in Appendix~\ref{sec:Pastur}.

\begin{figure}
\begin{center}
\includegraphics[width=0.48\textwidth]{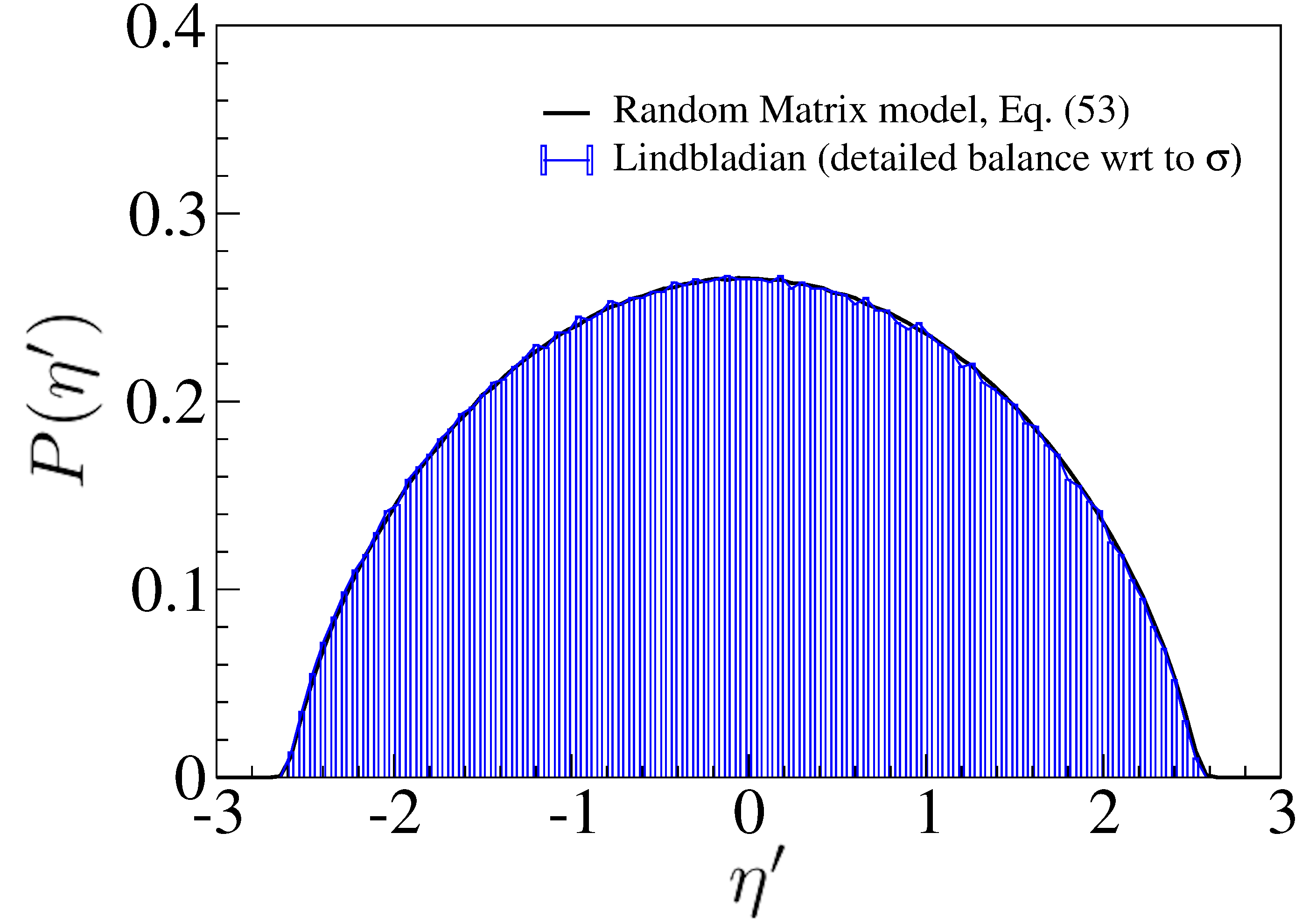}
\end{center}
\caption{Density of rescaled eigenvalues of random Lindblad operators obeying quantum detailed balance condition with respect to the maximally mixed state $\sigma =\oper/N$ (blue bars) and the density of the RM model, Eq.~\eqref{eq:Lindblad_max_mixed}, obtained by solving Eq.~\eqref{eq:Pastur_numeric} numerically (solid black line). Both densities for $N=100$ are sampled with $10^2$ realizations.
}
\end{figure}

\subsection{Detailed balance with respect to random stationary state with no degeneracy}
\label{sec:Davies_nondegen}

If the stationary state is chosen at random  with respect to any non-atomic measure \cite{ZPNC11}, its spectrum is non-degenerate with probability one. The corresponding Lindblad operator is characterized by Theorem~\ref{Thm:Lindblad_nondegen}.  The corresponding diagonal form reads

\begin{equation}\label{}
  \mathcal{L}_{\rm Dav}(\rho) = \sum_{i,j=1}^N L_{ij} \rho L_{ij}^\dagger - \frac 12 \{ L_{ij}^\dagger L_{ij} \rho \} ,
\end{equation}
where

\begin{itemize}
\item $L_{ii} = \sum_{j=1}^{N}l^{i}_{j}E_{jj}$, %for $i=1,\ldots,N-1$ and $l^{ii}_{jj}\in \mathbb{R}$ satisfying $\sum_{j=1}^{N} l^{ii}_{jj}=0$
\item $L_{ij} = S_{ij}p_j^{-1/2}E_{ij}$ for $i,j=1,\ldots,N$, $i\neq j$ and $S_{ij}\in \mathbb{C}$ such that $|S_{ij}| = |S_{ji}|$.
\end{itemize}
Recall that $E_{ij} =  \ket{i}\bra{j}$. $S_{ij}$ and $S_{ji}$ are related up to their phases and we choose $S$ Hermitian. Theorem~\ref{Thm:Lindblad_nondegen} is recovered by setting $D_{ij} = \sum_k l^k_i l^k_j$ and $W_{ij} = |S_{ij}|^2p_j^{-1}$. 

 By a direct inspection of the Lindblad operator \eqref{eq:Lindblad_nondegen} one finds

\begin{equation}\label{}
  \cL_{\rm Dav}(E_{ij}) = \lambda_{ij} E_{ij} \ \ (i\neq j) ; \ \qquad \cL_{\rm Dav}(E_{ii}) = \sum_{j=1}^{N} \cK_{ij} E_{jj} ,
\end{equation}
where real eigenvalues $\lambda_{ij}$ read $\lambda_{ij} = -\frac{1}{2}(d_{ij} + w_i + w_j)$, and
\begin{equation}\label{eq:DaviesEgv}
d_{ij}= \sum_{k=1}^{N}(l_{i}^{k} - l_{j}^{k})^2 ,\qquad w_i = \sum_{k\neq i} |S_{ki}|^2p_i^{-1},
\qquad \cK_{ij} = |S_{ij}|^2 p_j^{-1} - \delta_{ij}\sum_{k=1}^{N} |S_{kj}|^2 p_j^{-1}.
\end{equation}
This means that QDB implies that the evolution of diagonal elements of density matrix completely decouples from the off-diagonal elements. Hermiticity preservation ties the evolution of the elements on the opposite side of diagonal and implies double degeneracy of the corresponding eigenvalues, since $\lambda_{ij}=\lambda_{ji}$. Besides this, off-diagonal elements of the density matrix evolve independently. Therefore, the Davies generator in its vectorized form can be decomposed into 
\begin{equation}
\hat{\cL}_{\rm Dav} = \cK \oplus \cD \oplus \cD,
\end{equation}
with $\cK$ being a Kolmogorov operator satisfying the classical detailed balance w.r.t. the probability vector consisting of  eigenvalues of $\sigma$. The diagonal matrix $\cD$ of size $N(N-1)/2$ contains the decoherence eigenvalues $\lambda_{ij}$ for $1\leq i< j \leq N$. Interestingly, the same structure of the Lindblad operator is obtained a result of action of superdecoherence~\cite{PRE}. Indeed, random Davies generator is almost classical. Only $d_{ij}$ in Eq.~\eqref{eq:DaviesEgv} is purely quantum and does not survive superdecoherence.

We introduce randomness in the Lindblad operator as follows. Elements of diagonal jump operators $\{L_{ii}\}$ are Gaussian with zero mean and variance $\sigma^2_l = \<(l^{i}_{j})^2\> = \frac{1}{N^2}$, while the elements of a Hermitian matrix $S_{ij}$ defining the off-diagonal jump operators $\{L_{ij}\}$ are complex Gaussian with zero mean and variance $\sigma^2_S = \<|S_{ij}|^2\> = \frac{1}{N} \frac{1}{\<p^{-1}\>}$, where we introduced a short-hand notation

\begin{equation}\label{}
\<p^{-1}\> := \frac{1}{N}\sum_{j=1}^{N} p_j^{-1} .
\end{equation}
 This normalization scheme ensures that $\<\tr K\> = N$.
In the numerical setting, we further rescale all jump operators by the same factor to have strict equality $\tr K = N$.

Note that $\sum_{k\neq i} |S_{ki}|^2$ is a sum of squares of independent complex Gaussian random variables. Hence,  its distribution is the same as of $\frac{1}{2}\sigma_S^2 \chi^2_{2(N-1)}$, where $\chi^2_{k}$ denotes chi-squared distribution with $k$ degrees of freedom. By the law of large numbers, at large $k$ $\chi^2_k$ is well approximated by Gaussian distribution with mean $k$ and variance $2k$. Therefore, for large $N$ we have
\begin{equation}
w_i \approx (1+q_i)(1+\xi_i N^{-1/2})
\end{equation}
where $\xi_i$ is a random variable from a standard normal distribution and $q_i = \frac{p_i^{-1}}{\<p^{-1}\>}-1$.

By following a similar reasoning one can found that, in the large $N$ limit, $d_{ij}$ is well approximated by Gaussian distribution with mean $2/N$ and standard deviation of $2\sqrt{2} N^{-3/2}$. This means that $d_{ij} \ll w_i$ and hence the contribution of $d_{ij}$ to the eigenvalues $\lambda_{ij}$ is negligible.

The decoherence eigenvalues are effectively described as follows
\begin{equation}
\lambda_{ij} \approx -1 - \frac{q_i+q_j}{2} + \frac{1}{\sqrt{2N}} \xi_{ij} ,
\end{equation}
where $\xi_{ij}$ are i.i.d Gaussian random variables with zero mean and unit variance.

The elements $|S_{ij}|^2$ are independent with standard deviation $\sigma^2_S$. Therefore, the spectrum of a symmetric matrix with elements $|S_{ij}|^2$ fits the Winger semicircle rescaled by $\sqrt{N}\sigma_S^2$. The term $-p_j^{-1}\sum_{k=1}^{N} |S_{kj}|^2$ is the sum of independent random variables, thus, by the Central Limit Theorem, its elements are Gaussian with mean $\frac{-1}{\<p^{-1}\>}$ and standard deviation $\frac{1}{N^{1/2}\<p^{-1}\>}$. Finally, we introduce diagonal matrix $Q = \textup{Diag}[q_1,q_2,\ldots,q_N]$. The Kolmogorov operator can be then represented as
\begin{equation}
\cK \approx \left(-\oper_N + \frac{1}{\sqrt{N}}\left({\rm GOE}_N + {\rm Gauss}_N\right)\right)\left(1+Q\right),
\end{equation}
with a $N\times N$ GOE matrix and a diagonal matrix with elements drawn from a standard normal distribution.

\begin{figure}
\begin{center}
\includegraphics[width=0.8\textwidth]{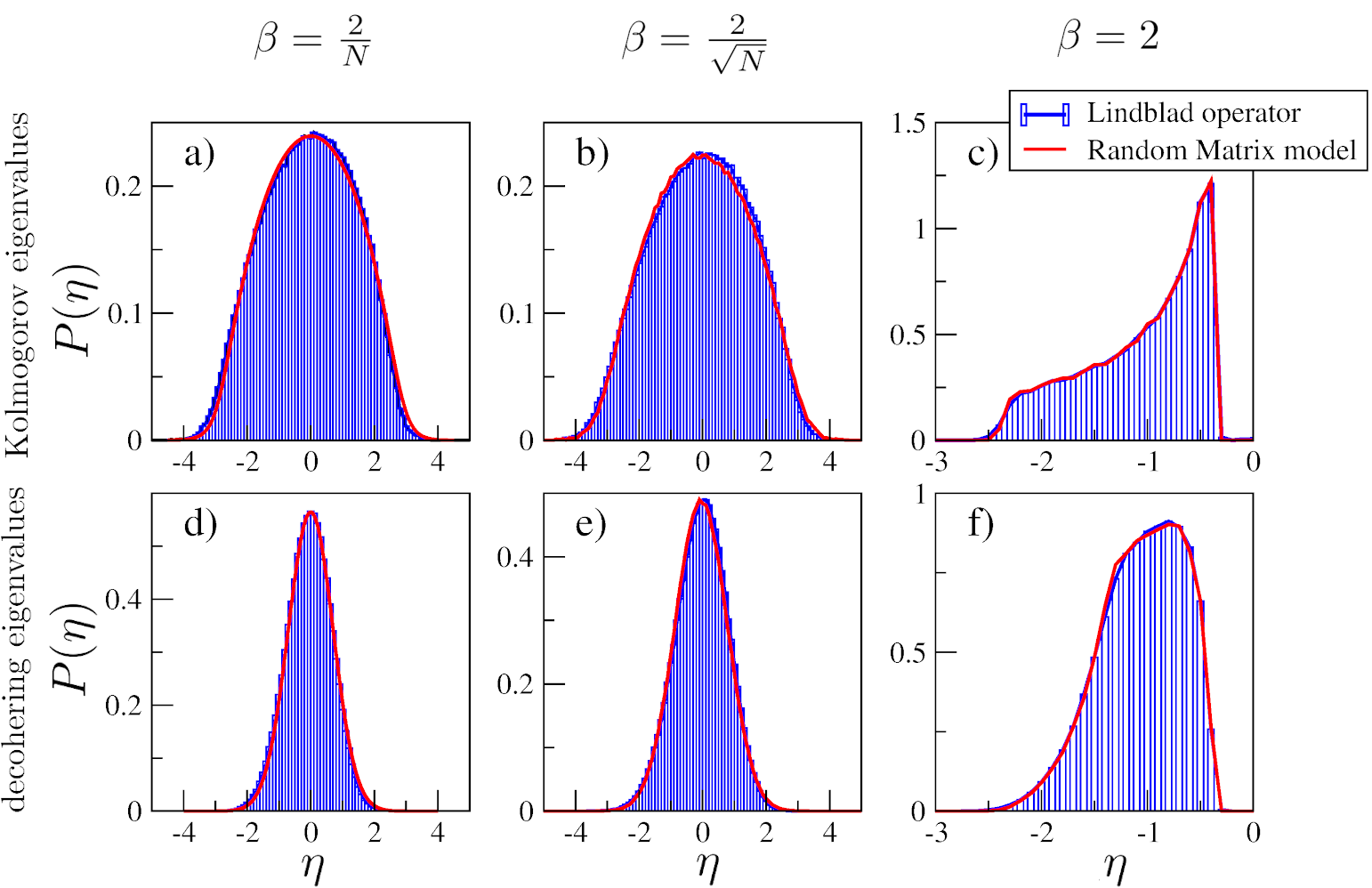}
\end{center}
\caption{Spectral  densities of random Lindblad operators satisfying quantum detailed balance condition  with respect to the thermal state, Eq.~\eqref{eq:ThermalState}, for three different regimes. Since decohering eigenvalues (bottom row) are doubly degenerated, they can be distinguished from the eigenvalues of the Kolmogorov part (top row) in the numerical procedure. In the first regime, $\beta \ll N^{-1/2}$, the density of the Kolmogorov part is juxtaposed with the spectral density of the RMT model described by Proposition~\ref{prop:GDavies}. In two other regimes densities of random Davies are compared to the spectral densities obtained by sampling  the corresponding RM models, Eqs.~\eqref{eq:ModelKolm2} and \eqref{eq:ModelKolm3}.
\label{fig:Davies}
}
\end{figure}

Note that there are two sources contributing to the randomness in random Davies operators: randomness in probabilities in $\sigma$ and randomness in the elements of the symmetric matrix $S$. We denote by $\kappa$ the standard deviation of $q_i$, which essentially encodes fluctuations in the probability vector $p$. It determines three regimes, each with different effective RM model.

\begin{enumerate}[label=(\roman*)]
\item If $\kappa \ll N^{-1/2}$, the randomness in $S$ dominates.  Kolmogorov and purely decohering components are rescaled as $\cK = -\oper_N + N^{-1/2} \cK'$ and $\cD = -\oper_{N(N-1)/2} + N^{-1/2}\cD'$, where
\begin{equation}
\cK' \approx {\rm GOE}_N + {\rm Gauss}_N,\qquad \cD' \approx \frac{1}{\sqrt{2}} {\rm Gauss}_{N(N-1)/2}. \label{eq:ModelKolm1}
\end{equation}

\item If $\kappa \gg N^{-1/2}$, the randomness in $\sigma$  dominates and
\begin{equation}
\cK_{ij} \approx - \delta_{ij} \frac{p_i^{-1}}{\<p^{-1}\>},\qquad \qquad \lambda_{ij} \approx -\frac{p_i^{-1} + p_j^{-1}}{2\<p^{-1}\>}. \label{eq:ModelKolm2}
\end{equation}

\item If $\kappa \sim N^{-1/2}$, both sources of randomness have comparable contributions that add up. Kolmogorov component and purely decohering eigenvalues are rescaled as $\cK = -\oper_N + N^{-1/2} \cK'$ and $\lambda_{ij} = -1 + N^{-1/2} \lambda'_{ij}$. With the introduction of rescaled variables $q_i = N^{-1/2}q'_i$ and $Q' = \textup{Diag}[q_1',\ldots,q_N']$, the rescaled components of the Davies operator read
\begin{equation}
\cK' \approx {\rm GOE}_N + {\rm Gauss}_N - Q', \qquad \lambda'_{ij} \approx -\frac{q'_i + q'_j}{2} + \frac{\xi_{ij}}{\sqrt{2}}, \label{eq:ModelKolm3}
\end{equation}
where $\xi_{ij}$ are standard normal variables independent for $i<j$ and satisfying $\xi_{ij}=\xi_{ji}$.

\end{enumerate}

While the densities in regimes (ii) and (iii) are dependent on a particular choice of the steady state probabilities, in regime (i), in the large $N$ limit, the corresponding spectral density of the Kolmogorov part can be calculated using the tools from free probability and its Stieltjes transform satisfies Pastur equation, see Appendix~\ref{sec:Pastur}.

\begin{Prop} \label{prop:GDavies}
Let $\cK' = A + B$, where $A$ is a GOE matrix normalized to $\<\frac{1}{N}\tr A^2\> = 1$ and $B$ is diagonal the elements of which are  independent Gaussians of mean zero and unit variance, as in Eq.~\eqref{eq:ModelKolm1}. The Stieltjes transform of the spectral density of $\cK'$ satisfies the functional equation
\begin{equation}
G(z) = G_{\cK}(z-G(z)),
\end{equation}
where $G_{\cK}$ is given by Eq. \eqref{eq:GKolmogorov}.
\end{Prop}

The above analysis can be illustrated  with a thermal state $\sigma_{\beta} = \mathrm{Diag}[p_1,p_2,...,p_N]$, where
\begin{equation}\label{eq:ThermalState}
  p_j = e^{- \beta E_j}/\mathcal{Z} \ , \ \ \ \mathcal{Z} = \sum_{j=1}^N e^{- \beta E_j}.
\end{equation}
If  energies $E_j$ are uniformly distributed on $[0,1]$, then the three regimes determined by tightness of distribution of $p_j^{-1}$ around its mean can be translated to the scaling of the inverse temperature: $\beta \ll N^{-1/2}$, $\beta \sim N^{-1/2}$, and $\beta \gg N^{-1/2}$. Hence for each regime we apply a different RM model (red line).

In Figure~\ref{fig:Davies}  we compare the numerical densities obtained by sampling over random Lindblad operators and spectral densities of the corresponding RM models, confirming validity of RM models in all 3 regimes.

\subsection{Partially degenerate stationary state}

Noting the remarkable difference in the spectra of Lindblad operators in the two scenarios, of the asymptotic state as the fully degenerate density matrix and density matrix with no degenerate eigenvalues, here we consider Lindblad operators interpolating between these two extreme cases. To this end, we consider density matrices in which first $M$ ($1\leq M \leq N$) eigenvalues are equal to each other and the remaining $N-M$ are pairwise different. For convenience, we denote by $A$ the set of indices in the degenerate space of $\sigma$ and by $p_A$ the corresponding probability.  We also assume that there are no additional relations between eigenvalues that would allow for additional non-zero elements of the Kossakowski matrix. Non-zero elements of the Kossakowski matrix are characterized as follows.

\begin{Prop} \label{prop:Kdegen}
Let $A =  \{1,\ldots,M\}$ and $p_i$ be the eigenvalues of $\sigma$ that satisfy the following relations:
\begin{enumerate}[label=(\roman*)]
\item $p_i=p_j$ if $i,j\in A$
\item $p_i\neq p_j$ if $i\notin A$ or $j\notin A$
\item $p_i p_j \neq p_{i'} p_{j'}$ and if one of indices $\notin A$, then $\{i,i'\}\neq \{j,j'\}$
\end{enumerate}
The non-zero elements of $K$ satisfying QDB are of the form:
\begin{enumerate}[label=(\roman*)]
\item $K_{ij,kl}$ for $i,j,k,l \in A$ satisfying $K_{ij,kl} = K_{lk,ji}$,
\item $K_{kk,ll}$ for $k,l\notin A$ satisfying $K_{kk,ll} = K_{ll,kk}$,
\item $K_{ij,kk}$ and $K_{kk,ji}$ for $i,j\in A$ and $k\notin A$ satisfying $K_{ij,kk}=K_{kk,ji}$,
\item $K_{ik,jk}$ and $K_{kj,ki}$ for $i,j \in A$ and $k \notin A$ satisfying $K_{ik,jk}p_k = K_{kj,ki}p_j$,
\item $K_{kl,kl}$ for $k,l \notin A$ satisfying $K_{kl,kl}p_l = K_{lk,lk}p_k$.
\end{enumerate}
\end{Prop}

The proof simply follows from the consistency condition $K_{ij,kl} = \frac{p_j p_k}{p_i p_l}K_{ij,kl}$ (cf.  \eqref{CC-2}), i.e., either $p_jp_k=p_ip_l$ or the corresponding element $K_{ij,kl}$ vanishes. The relation between those elements are determined by QDB, Eq.~\eqref{CC}.

By reordering indices of $K$, we bring it into a block-diagonal form with: One block of size $M^2+N-M$ containing cases (i)-(iii) in Proposition~\ref{prop:Kdegen}, $2(N-M)$ blocks of size $M$ (elements are indexed by $i$ and $j$) corresponding to case (iv), and a diagonal part of size $(N-M)(N-M-1)$ containing elements described by the case (v). Note that in cases (i) - (iii) QDB implies equality of certain elements within the block, while for cases (iv) - (v) it imposes relations between certain pairs of blocks.

Having identified the block structure of $K$ and knowing that it is positive definite, we represent it as $K=G^{\dagger}G$ and choose $G$ that has the same block structure as $K$. Then, the jump operators are constructed in a similar way as in Eq.~\eqref{eq:JumpLemon}, but this time with the basis matrices $E_{ij}$ and we immediately arrive at the following

\begin{Prop} \label{prop:jump_part_degen}
Let the Lindblad generator be given by the following set of jump operators:
 \begin{enumerate}[label=(\roman*)]
\item $L_{\alpha} = \sum_{i,j=1}^{M}l^{\alpha}_{ij} E_{ij} + \sum_{k=M+1}^{N}l^{\alpha}_{kk}E_{kk}$ with $l^{\alpha}_{ij} = \overline{l}^{\alpha}_{ji}$ for  $\alpha = 1, \ldots,  M^2+N-M$,
\item $L_{kl} = S_{kl} p_{l}^{-1/2} E_{kl}$ for $k,l=M+1,\ldots, N$ and $k\neq l$,
\item $L_{ik} = \sum_{j=1}^{M} T^{i}_{jk}p_k^{-1/2} E_{jk}$ for $i=1,\ldots,M$ and $k=M+1,\ldots,N$,
\item $L_{ki} = \sum_{j=1}^{M} T^{i}_{kj}p_{j}^{-1/2} E_{kj}$ for $i=1,\ldots,M$ and $k=M+1,\ldots,N$,
\end{enumerate}
where complex matrices $S$ and $T^i$ satisfy $|S_{kl}|=|S_{lk}|$ and $T^{i}_{jk} = \overline{T^{i}_{kj}}$. Then this generator satisfies the QDB condition with respect to $\sigma$ whose eigenvalues satisfy assumptions of Proposition \ref{prop:Kdegen}.
\end{Prop}

\begin{Remark}
Davies generator is recovered at $M=1$ where cases (iii) and (iv) reduce to (ii), or, eqivalently, at $M=0$ where (iii) and (iv) are empty. In case of full degeneracy, all jump operators are of the form (i). The presence of non-trivial generators of types (iii) and (iv) is a unique feature of QDB Lindblad operators with partially degenerate steady state.
\end{Remark}

\begin{figure}
\begin{center}
\includegraphics[width=\textwidth]{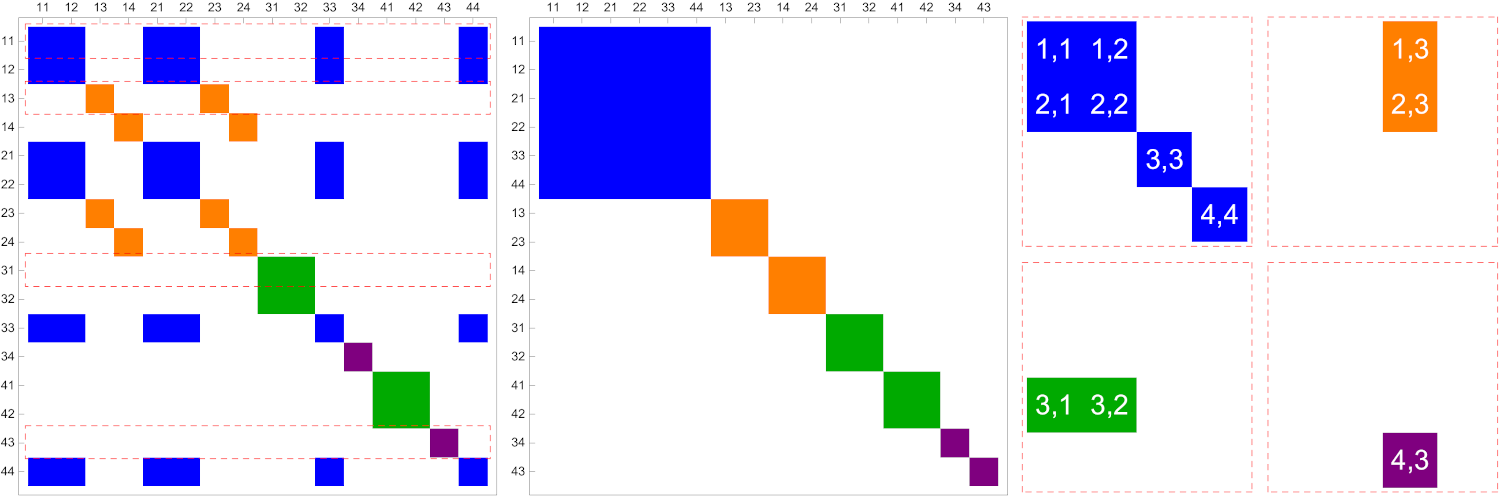}
\end{center}
\caption{(left) Kossakowski matrix satisfying quantum detailed balance condition for a system of size $N=4$ with $M=2$ degenerated eigenvalues. (center) Upon permutation of its indices, the Kossakowski matrix can be brought into a block-diagonal form. (right) Non-zero elements of $K$ indicate non-zero elements in jump operators, which are obtained by taking a row of $K$, slicing it into pieces of length $N$ each. These pieces are stacked   on top of one another to form $L$.
\label{fig:KossakConstr}
}
\end{figure}

Elements $S_{ij}$ and $S_{ji}$ have equal modulus, but their phases are not constrained. For convenience, we choose $S$ Hermitian.
We also introduced a hybrid notation in which some jump operators are index with a single Greek letter, while others with double-index Latin letter. This construction of jump operators is illustrated in Fig.~\ref{fig:KossakConstr}. The Lindblad operator has the same block structure as the underlying Kossakowski matrix, see Appendix~\ref{sec:LindbladSpectrum}.

As observed in Section~\ref{sec:Davies_nondegen}, the shape of the spectrum of random Davies generators depends on the scale of randomness in the matrix $S$ and the spread of inverse eigenvalues of the steady state. The motivation for the interpolating Lindblad operators was to study the influence of the degree of the degeneracy and the resulting structure of the Kossakowski matrix on the shape of the spectrum. From that perspective, distribution of $p_a^{-1}$ may blur the resulting picture by bringing unnecessary complexity. Therefore, we assume here that first $M$ eigenvalues of $\sigma$ are equal to $1/N$, while the non-degenerate eigenvalues are  concentrated around $1/N$. The spectrum of $\cL$ is not affected by the distribution of $p_a$, so we can set $p_a=1/N$.

We introduce randomness into jump operators in the following way. Elements of complex hermitian matrices are Gaussian with zero mean and variances $\<|l^{\alpha}_{ij}|^2\>=\sigma^2_L$, $\<|T^{i}_{jk}|^2\>=\sigma^2_T$ and $\<|S_{ij}|^2\>=\sigma^2_S$. We assume the following normalization
\begin{align*}
\sigma_L^2 = \frac{1}{N(M^2-M+N)},\qquad \sigma_S^2 = \frac{1}{N^2},\qquad \sigma^2_T = \frac{1}{MN^2}.
\end{align*}
This normalization scheme assures that jump operators have on average equal norm,  $\<\tr LL^{\dagger}\> = \frac{1}{N}$.

The spectra of Lindblad operators are centered at $-1$ and the eigenvalues are scattered in an interval the size of which scales like $\frac{\sqrt{N-M+1}}{N}$. This scaling factor interpolates between $N^{-1/2}$ for random Davies generators ($M=1$) and $N^{-1}$ for the case of maximimally mixed stationary state. Therefore, to describe bulk of the spectrum we rescale the Lindblad operator as $\cL = -\oper + \frac{\sqrt{N-M+1}}{N} \cL'$, and propose a random matrix model valid for $1\ll M \ll N$.
The random matrix model is composed of four components, $\cL' \approx \cL'_1 \oplus \cL'_1 \oplus \cL'_2 \oplus \cL'_2 \oplus \cL'_3 \oplus \cL'_4$, with the building blocks
\begin{align}
\cL'_1 \approx   \frac{1}{\sqrt{2}} \textup{Gauss}_{(N-M)(N-M-1)/2}, &\qquad  \cL'_2 \approx    \bigoplus_{i=1}^{N-M} \frac{1}{2}(\textup{Gauss}_M^{(i)} + \textup{GOE}_{M}^{(i)}),
\\
\cL_3' \approx \textup{GOE}_{M} \otimes \oper_M + \oper_M \otimes \textup{GOE}_{M}, & \qquad   \cL'_4 \approx \textup{GOE}_{N-M} + \textup{Gauss}_{N-M}. \label{eq:RMMpartdegen}
\end{align}
Here $\textup{Gauss}$ is a diagonal matrix with standard independent Gaussian elements on the diagonal. Blocks $\cL'_1$ and $\cL'_2$ appear twice to reflect the double degeneracy of eigenvalues due to Hermiticity preservation. Models for $\cL'_3$ and $\cL'_4$ are found under a further approximation of the part of Lindbladian given by jump operators of type (i) in Proposition~\ref{prop:jump_part_degen}, see Appendix~\ref{sec:Justification} for details.

With the decreasing degeneracy, blocks $\cL_2'$ and $\cL'_3$ shrink and disappear for $M=0$, while $\cL_1'$ and $\cL'_4$ reduce to the random matrix model for random Davies as in Eq. \eqref{eq:ModelKolm1}. The route to the matrix model for a maximally mixed state is less trivial. The RMM for $\cL'_3$ should include also the term $\frac{1}{\sqrt{N-M+1}}\cL_{\rm mm}'$, however, since we assume $N\gg M$ this part is neglected. On the other hand,  for $M \approx N$, the GOE matrix in $\cL'_3$ ceases to be a good model and this regime requires more refined treatment, see  Appendix~\ref{sec:Justification} for details.

\begin{figure}
\begin{center}
\includegraphics[width=0.65\textwidth]{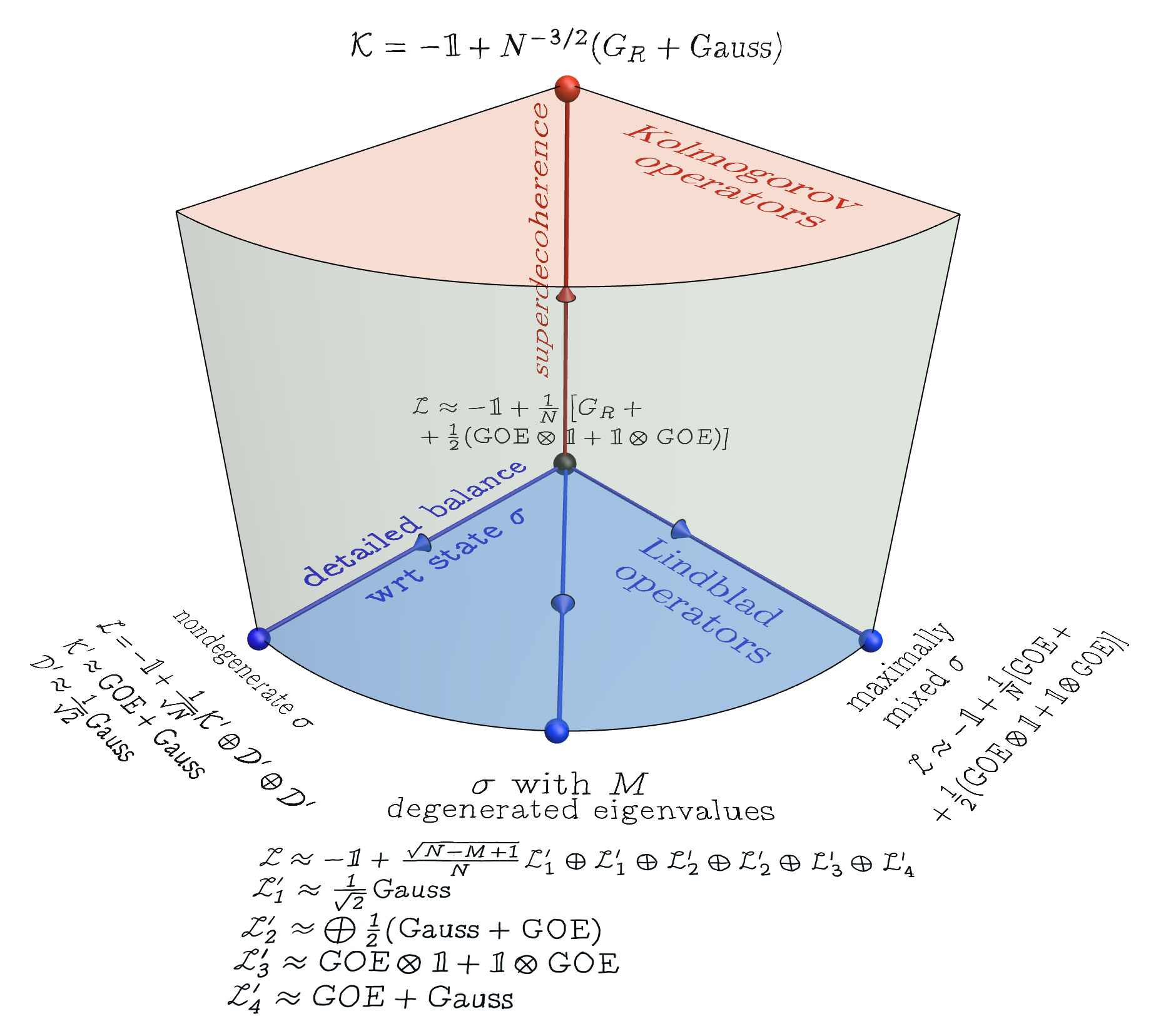}
\end{center}
\caption{Ensembles of random generators of Markovian evolution obeying detailed balance and the corresponding Random Matrix models.
Quantum generators (Lindblad operators) are specified by the type of their asymptotic states.  For any ensemble of quantum Lindblad generators $\cal L$
one obtains, by applying supredecoherence,
the corresponding ensemble of classical  Kolmogorov operators $\cal K$.}
\end{figure}

\section{Conclusions}
\label{sec:Conclusions}

We introduced several ensembles of random Lindblad operators $\cal L$,
which satisfy the quantum detailed balance condition.
These ensembles
can be labeled with their stationary state $\sigma$.
Random stationary states, which are typically non-degenerate,
lead to the so-called {\sl Davies generators}~\cite{Davies1,Davies2,Davies3}.
As the detailed balance condition
enforces the spectrum of $\cal L$ to be real,
we investigated   density of eigenvalues along the real axis.
For various assumptions concerning the degeneracy of  stationary state $\sigma$,
we constructed the corresponding ensemble of random matrices (see Appendix \ref{sec:Glossary}),
which allowed us to find analytic expressions for the asymptotic spectral densities.

In our previous works~\cite{Lemon,PRE}, we explored properties
of completely random Lindblad operators and found that their spectra are supported
on a universal lemon-shaped region on the complex plane. Integrating out the imaginary
part of the eigenvalues, we obtain the marginal distribution which provides a 
fair approximation to the spectral density of Lindblad generators obeying detailed balance
with respect to the maximally mixed state.

We also consider random Kolmogorov operators, which induce time-continuous dynamics over the classical probability simplex. Instead of analyzing
the problem in a straightforward manner,  we employed the idea of superdecoherence~\cite{PRE}. Namely, by decoherefying  ensembles of quantum Davies generators, we transform them into classical Davies generators 
and investigated spectral features of the latter. As in the quantum case,
the marginal of the probability distribution describing the
complex spectra of random Kolmogorov operators ${\cal K}$,
supported on the universal spindle-shaped region on the complex plane \cite{Timm1, BordenaveChafai},
gives a reasonable approximation of the  spectral density of classical Davies generators.

Finally, we developed a family of Random Matrix models which reproduce spectral densities of different balance-obeying operators, classical and quantum.
They are summarized on a sketch presented in Figure 5.

As a next step, it would be interesting to investigate  how the  quantum-to-classical transition, induced by superdecoherence,
realizes in case of generators satisfying (exactly on in part) the detailed balance condition. Furthermore, the spectral properties
of random Lindblad operators depend significantly on the rank of the operator,
which could take any value from $1$ to $N^2$.
A more exhaustive analysis of random generators
and their dependence on three parameters,
(a) degree of superdecoherence,
(b) accuracy with which the detailed balance condition is satisfied, and
(c) rank of operator $\cal L$,
will be a subject of a forthcoming work \cite{T+23}.

\section{Acknowledgments}
\label{sec:Acknowledgment}
This research is supported by Research Council of Norway,
project ``IKTPLUSS-IKT og digital innovasjon - 333979" (SD) and by 
National Research Center (Poland),
project 2021/03/Y/ST2/00193
(WT and K\.Z), as parts of the ERA-NET project ``DQUANT: A Dissipative Quantum Chaos perspective on Near-Term Quantum Computing". DC was supported by the Polish National Science Center project No. 2018/30/A/ST2/00837.

\section*{Appendices}
\appendix

\section{Basic ensembles of random matrices}
\label{sec:Glossary}

  The following fundamental ensembles of
random matrices were used in this paper:
\bigskip
\begin{itemize}

\item  Let $G$ denote a random $N\times N$ {\bf Ginibre matrix} \cite{Gi65},
    with the entries being i.i.d. Gaussian random
    variables  with zero mean and variance
     $\langle |G_{ij}|^2 \rangle=1/N$.
    Its spectrum covers uniformly the unit disk, a result called the \emph{circular
	law of Girko} \cite{Gi65, Gi84,Fo10}.

\item
    $G_R=(G+ \overline{G})/\sqrt{2}$ denotes a matrix from the {\bf real Ginibre matrix} - real non-symmetric
   random  matrix. The spectrum consists of a (deformed) Girko disk in a complex plane
    and a singular component at the real axis, compensated by a dip in the spectral density
    just below and above the real axis. In the limit $N\to \infty$ these finite--size effects
      disappear~\cite{LS91,AkemannKanzieper,ForresterNagao}, and the standard Girko disk is recovered.

      \item
       $\textup{GUE}_N = (G+ G^{\dagger})/\sqrt{2}$ denotes
          a hermitian random matrix from
      {\bf Gaussian Unitary ensemble} \cite{Me04}. The name refers
        to the invariance of its probability density with respect to the unitary group $U(N)$.
         In the limit of large matrix dimension $N$,
	its spectrum converges to Wigner's
	semicircle distribution~\eqref{Wigner-C}.

  \item
       ${\rm GOE}_N=({\rm GUE}_N+ \overline{\rm GUE}_N)/\sqrt{2}$ denotes
          a real symmetric random matrix from
      {\bf Gaussian Orthogonal ensemble},
         invariant with respect to the orthogonal  group $O(N)$.
        Asymptotically the  level density also converges to the semicircle,
          but the correlation between levels are different.
              The nearest-neighbour distribution displays level repulsion,
            $P(s)\sim s^{\beta}$, with $\beta=2$ for GUE and
            $\beta=1$ for GOE, where $s$ denotes
            the (unfolded) spacing between consecutive eigenvalues.

   \item $\textup{Gauss}_N$ denotes a random diagonal matrix
               with independent real Gaussians with zero mean and unit variance at the diagonal.
                 Its level density is Gaussian, and the
               uncorrelated levels exhibit {\sl level clustering} and Poissonian behaviour,
           $P(s) = \exp(-s)$, which corresponds to $\beta=0$.
   \item Finally,  $X$ is an auxiliary $N\times M$ real rectangular Ginibre matrix. Its elements are real i.i.d Gaussian random variables  with zero mean     	and unit variance. The $N\times N$ matrix $\textup{Wish}_{N,M} = \frac{1}{M}XX^T$ is a real \textbf{Wishart} matrix. In the limit                       	$N,M\to\infty$ such that $r=N/M$ remains constant its spectral density is given by the Marchenko-Pastur law~\cite{MarchenkoPastur,PasturBook}
   \begin{equation}
   \rho_{MP}(x) = \frac{1}{2\pi r x}\sqrt{(r_{+}-x)(x-r_{-})},
   \end{equation}
   where $r_{\pm} = (1\pm \sqrt{r})^2$ are the edges of the spectrum. If $r>1$ the density includes additional Dirac delta at $0$.

 \end{itemize}

\section{Pastur equation}
\label{sec:Pastur}

The main object of interest in Hermitian random matrices is the average spectral density $\rho(x) = \<\frac{1}{N}\sum_{i=1}^{N}\delta(x-\lambda_i)\>$. This object is studied through its Stieltjes transform, also known in the physics literature as the Green's function, $G_{H}(z) = \< \frac{1}{N} \tr (z-H)^{-1}\>$. Once the complex-valued Green's function is known, the spectral density is recovered from the behavior of $G$ near real line via Sochocki-Plemelj formula
\begin{equation}
\rho(x) = \mp \frac{1}{\pi}\lim_{\epsilon \to 0} G(x\pm i\epsilon).
\end{equation}

One of the central problems in random matrix theory is to calculate the spectrum of a sum of two random matrices $A+B$. This problem can be solved in the large $N$ limit with the tools of free probability developed for non-commuting objects~\cite{VND,MSp17}. In this framework the notion of independence is replaced by freeness. Informally, two matrices are free if there are no relations between their eigenbases.

To calculate the spectrum of a sum of two matrices, one first calculates their Green's function and its function inverse $B(z)$ satisfying
\begin{equation}
G(B(z)) =z,\qquad B(G(z))=z, \label{eq:GBrelation}
\end{equation}
and defines $R(z) = B(z) -\frac{1}{z}$ the $R$-transform, which is additive for free random matrices
\begin{equation}
R_{A+B}(z) = R_A(z) + R_B(z).
\end{equation}

In the context of this work we are interested in the case when $A$ is a GOE, while $B$ can be arbitrary Hermitian. The $R$-transform for GOE reads $R(z)=z$, thus we have the relation $R_{A+B}(z) = z+ R_{B}(z)$. Adding $\frac{1}{z}$ to both sides leads to
\begin{equation}
B_{B}(z) = B_{A+B} - z.
\end{equation}
In the net step we evaluate $G_B$ at both sides of equation and use \eqref{eq:GBrelation} to get
\begin{equation}
z = G_B(B_{A+B}(z)-z).
\end{equation}
Finally, we substitute $z\to G_{A+B}(z)$ and use \eqref{eq:GBrelation}, obtaining the Pastur equation~\cite{PasturEquation}:
\begin{equation}
G_{A+B}(z) = G_{B}(z-G_{A+B}(z)).
\end{equation}

To cover cases appearing in the main text, one only needs to find $G_{B}(z)$. In fact, for both instances of our interest, namely $B=\frac{1}{2}(C \otimes \oper + \oper \otimes C)$ with $C$ a GOE matrix (for Proposition~\ref{Prop:GMaxMixed}) and $B$ Gaussian diagonal (for Proposition~\ref{prop:GDavies}), their Stieltjes transforms are calculated in~\cite{Lemon, PRE}.

\section{Spectral properties of Lindblad operators obeying detailed balance with respect to a partially degenerate steady state}
\label{sec:LindbladSpectrum}

By straightforward, albeit lengthy, calculation one can verify action of the QDB generators acting on the base matrices.

\begin{Prop} \label{Prop:LindblPartDegen}
The Lindblad operator be defined by the set of jump operators  in Proposition~\ref{prop:jump_part_degen} satisfies the following relations:
\begin{enumerate}[label=(\roman*)]

\item for $a,b\notin A$ and $a\neq b$
\begin{displaymath}
\cL(E_{ab}) =  \lambda_{ab} E_{ab},\quad \mbox{where}\quad  \lambda_{ab} = -\frac{1}{2}(d_{ab} + t_a + w_a + w_b + t_b),
\end{displaymath}

\item for $a\in A$ and $b\notin A$
\begin{align*}
\cL(E_{ab}) = -\frac{1}{2}\sum_{i=1}^{M}\left( G^{(b)}_{ia} +p_A^{-1} H_{ia} + \delta_{ia}(w_a+t_a)\right)E_{ib},
\end{align*}

\item for $a\notin A$ and $b\in A$
\begin{align*}
\cL(E_{ab}) = -\frac{1}{2}\sum_{i=1}^{M}\left( G^{(a)}_{bi} + p_A^{-1} H_{bi} + \delta_{bi}(w_b+t_b)\right)E_{ai},
\end{align*}

\item for $a,b\in A$
\begin{align*}
\cL(E_{ab}) = \widetilde{\cL}_{\rm mm}(E_{ab}) +p_{A}^{-1}\sum_{k=M+1}^{N}  \cT_{k,ba}E_{kk} - \frac{1}{2}p_{A}^{-1} \sum_{j=1}^{M}\left( H_{ja}E_{jb} + H_{bj}E_{aj}\right),
\end{align*}

\item for $a\notin A$
\begin{align*}
\cL(E_{aa}) = \sum_{k=M+1}^{N} \cK_{ka}E_{kk} + \sum_{i,j=1}^{M} p_a^{-1}\cT_{a,ij}E_{ij} - t_a E_{aa},
\end{align*}
\end{enumerate}
where
\begin{align*}
\cT_{a,ij} = \sum_{m=1}^{M} T_{ia}^m T^m_{aj},\qquad H_{ja} = \sum_{k=M+1}^{N} \cT_{k,ja},\qquad t_a = p_a^{-1}\sum_{j=1}^{M} \cT_{a,jj},  \qquad
w_a = \sum_{\substack{k=M+1 \\ k\neq a}}^{N} |S_{ka}|^2 p_a^{-1},
\end{align*}
\begin{align*}
d_{ab} = \sum_{\alpha} \left(l^{\alpha}_{aa}-l^{\alpha}_{bb}\right)^2,\qquad G^{(b)} = \sum_{\alpha} (J^{(b)}_{\alpha})^2,\qquad (J_{\alpha}^{(b)})_{ij} = l^{\alpha}_{ij} - l_{bb}^{\alpha}\delta_{ij},\qquad \widetilde{\cK}_{ka} = |S_{ka}|^2p_a^{-1} - \delta_{ka}\sum_{j=M+1}^{N} |S_{ja}|^2p_a^{-1},
\end{align*}
\begin{align*}
\widetilde{\cL}_{\rm mm}(E_{ab}) = \sum_{\alpha} \tilde{L}_{\alpha}E_{ab}\tilde{L}_{\alpha}-\frac{1}{2} \{\tilde{L}_{\alpha}^2,E_{ab}\} \qquad  \mbox{with} \qquad
\tilde{L}_{\alpha} = \sum_{i,j=1}^{M} l^{\alpha}_{ij} E_{ij}.
\end{align*}

\end{Prop}

The relation between (ii) and (iii) simply follows from hermiticity preservation of $\cL$. $\widetilde{\cL}_{\rm mm}$ above is a Lindbladian satisfying QDB with respect to a maximally mixed state, restricted to the space spanned by the degenerate part of $\sigma$. Analogously, $\widetilde{\cK}$ is a Kolmogorov generator restricted to the space spanned by non-degenerate part of $\sigma$.

The proposition above shows that $\hat{\cL}$ takes a block-diagonal form directly coinciding with the structure of the Kossakowski matrix.
The Lindbladian is already diagonal with eigenvalues $\lambda_{ab}$ on the subspace spanned by $E_{ab}$ with $a\neq b$ corresponding to the non-degenerate part of $\sigma$. There are $(N-M)(N-M-1)$ such eigenvalues including their double degeneracy. Moreover, there are $N-M$ blocks (each for $b\notin A$) of size $M\times M$. The corresponding eigenvalues are doubly degenerate as well, since cases (ii) and (iii) are governed by the same matrix, up to a transposition. Finally, there is a block of size $M^2-M+N$ governing dynamics of diagonal elements of the density matrix and the elements in the subspace spanned by degenerate part of $\sigma$.

\section{Justification of the random matrix model}
\label{sec:Justification}

In this section we analyze blocks of the random Lindbladian defined as in Proposition~\ref{Prop:LindblPartDegen} and justify the random matrix model~\eqref{eq:RMMpartdegen}.

\textit{Case (i).} All eigenvalues are of the form $\lambda_{ab} = -\frac{1}{2}(d_{ab} + w_a + w_b + t_a + t_b)$. Similarly as in the case of random Davies, contribution from $d_{ab}$ is negligible. Both $w_a$ and $t_a$ are given as sums of squared moduli of complex Gaussian, thus have the $\chi^2$ distribution. Overall, $\lambda_{ab}$ has the distribution as the random variable
\begin{equation}
\lambda_{ab} \sim -\frac{1}{4MN} \chi^2_{4M^2} - \frac{1}{4N} \chi^2_{4(N-M-1)}.
\end{equation}
For small $M$ the first term can be neglected, while for $N\gg M \gg 1$ both terms contribute. By the central limit theorem, $\chi^2_N$ is well approximated by the normal distribution with mean $N$ and variance $2N$. Therefore, for $M\ll N$ $\lambda_{ab}$ is well approximated by Gaussian distribution with mean $-1$ and variance $\frac{N-M}{2N^2}$.

\textit{Case (ii).} The elements of the matrix $J_{\alpha}^{(b)}$ are Gaussian, so we use Wick's theorem for the calculation of its moments. Its second moment reads $\<\frac{1}{M}\tr (J_{\alpha}^{(b)})^2\> = \sigma_L^2(M+1)$ and the fourth moment $\<\frac{1}{M}\tr (J_{\alpha}^{(b)})^4\> = \sigma_L^4(2M^2+6M+4)$. Therefore, the spectrum of the matrix $(J_{\alpha}^{(b)})^2$ has mean $\sigma_L^2(M+1)$ and variance $\sigma_L^4 (M^2+4M+3)$. The matrix $G^{(b)}$ is a sum of $M^2-M+N$ such matrices, therefore according to free CLT its spectrum is the Wigner semicircle centered at $M/N$ with variance $\frac{M^2}{N^2(M^2-M+N)}$. The variance is of order at most $1/N$ and decreases are $M$ grows. Therefore, it is negligible, when compared to the contribution from $t_a$ and $w_a$ analyzed before, which brings us to conclusion that $G^{(b)}$ can be approximated by the identity matrix rescaled by $\frac{M}{N}$.

The elements of matrix $H$ are given by
\begin{equation}
H_{ia} = \sum_{k=M+1}^{N} \sum_{m=1}^{M} T_{ik}^{m} T_{ka}^{m}. \label{eq:HWishart}
\end{equation}
The fact that $H$ is a Wishart matrix is the most evident if one notices that indices $k$ and $m$ can be grouped into a single index ranging from $1$ to $M(N-M)$. Therefore, $p_A^{-1}H$ has the same distribution as the matrix $\frac{N-M}{N} \textup{Wish}_{M,M(N-M)}$, where  $\textup{Wish}_{N,T}$ stands for $N\times N$ Wishart matrix with the rectangularity parameter $\frac{N}{T}$.

Putting these results together, the Lindbladian in a single $M\times M$ block of case (ii) can be represented as
\begin{align*}
-\frac{1}{2}\left( \frac{M+N}{N} \oper_{M} + \frac{\sqrt{N-M}}{N}\textup{Gauss}_M + \frac{N-M}{N} \textup{Wish}_{M,M(N-M)}\right).
\end{align*}
This model can be further simplified when $N-M$ is large, because the Wishart matrix can be approximated as
\begin{equation}
\label{eq:WishartSimplification}
\textup{Wish}_{M,M(N-M)}\approx \oper_M + \frac{1}{\sqrt{N-M}} \textup{GOE}_M.
\end{equation}
 This approximation follows from \eqref{eq:HWishart}, in which $H$ can be interpreted as a sum of $N-M$ independent $\textup{Wish}_{M,M}$ matrices to which free CLT is then applied. Therefore, the final model for a single $M\times M$ block of case (ii) reads
\begin{displaymath}
 -\oper_M + \frac{\sqrt{N-M}}{2N} (\textup{GOE}_{M} + \textup{Gauss}_M).
\end{displaymath}

\textit{Case (iii)} is a direct copy of case (ii) due to hermiticity preservation.

 According to Proposition~\ref{Prop:LindblPartDegen}, the Lindbladian couples elements $E_{aa}$ for $a\notin A$ and $E_{ij}$ for $i,j\in A$. However, following numerical results, we make the assumption of neglecting the mixing terms containing $\cT_{a,ij}$ in both cases, which gives rise to two separate blocks.

\textit{Case (iv).} The matrix model for $\hat{\cL}_{\rm mm}$ analyzed in section~\ref{sec:Davies_degen} needs to be adjusted to take into account different variance in the elements and different size. After these simple adjustments we have
\begin{equation}
\widetilde{\hat{\cL}}_{\rm mm} = - \frac{M}{N} \oper + \frac{1}{N} \hat{\cL}'_{\rm mm},
\end{equation}
where $\hat{\cL}'_{\rm mm}$ is given by \eqref{eq:Lindblad_max_mixed}. As argued in case (ii), the part with the $H$ matrix leads to
\begin{align*}
-\frac{N-M}{2N}\left( \textup{Wish}_{M,M(N-M)} \otimes \oper_M + \oper_M \otimes \textup{Wish}_{M,M(N-M)}\right),
\end{align*}
which can be further simplified using \eqref{eq:WishartSimplification}. Additionally, under this assumption the term with $\hat{\cL}'_{\rm mm}$ is subleading and we neglect it to get the final model
\begin{displaymath}
\hat{\cL}_3 \approx -\oper_{M^2} + \frac{\sqrt{N-M}}{2N}\left( \textup{GOE}_{M}\otimes \oper_{M} + \oper_M \otimes \textup{GOE}_M \right).
\end{displaymath}
For $N-M={\cal O}(1)$ the approximation of a Wishart matrix by GOE ceases to hold, therefore to capture the transition between partially degenerated case and fully degenerated case, one needs to use the full model
\begin{equation}
 \hat{\cL}_3 \approx -\frac{M}{N} \oper -\frac{N-M}{2N}\left( \textup{Wish}_{M,M(N-M)} \otimes \oper_M + \oper_M \otimes \textup{Wish}_{M,M(N-M)}\right) + \frac{1}{N} \hat{\cL}'_{\rm mm}.
\end{equation}

\textit{Case (v).} The first component is the truncation of the Kolmogorov generator, the elements of which are generated as in Section~\ref{sec:Davies_nondegen}, but only lower-left block of size $N-M$ is taken. Taking this into account,  the random matrix model is qualitatively  the same with a rescaling and shifting and reads
\begin{displaymath}
\widetilde{\cK} \approx -\frac{N-M}{N} \oper_{N-M} + \frac{\sqrt{N-M}}{N} \left(\textup{GOE}_{N-M} + \textup{Gauss}_{N-M}\right).
\end{displaymath}
As argued previously, $t_a$ has the distribution of $\frac{1}{2MN}\chi^2_{2M^2}$, which can be approximated by the normal distribution with mean $\frac{M}{N}$ and variance $1/N^2$. For $M\ll N$ the fluctuations are negligible as compared to $\frac{N-M}{N^2}$ above, therefore terms with $t_a$ can be approximated by $\frac{M}{N}\oper $, hence
\begin{equation}
\hat{\cL}_4 \approx - \oper_{N-M} + \frac{\sqrt{N-M}}{N} \left(\textup{GOE}_{N-M} + \textup{Gauss}_{N-M}\right).
\end{equation}

\end{document}